\documentclass{aa}  

\usepackage{txfonts}
\usepackage[colorlinks=true,linkcolor=blue,citecolor=blue,urlcolor=blue]{hyperref}
\usepackage{orcidlink}
\usepackage{adjustbox}
\usepackage[flushleft]{threeparttable}
\usepackage{comment}
\newcommand{\kms}{\ensuremath{\rm km\ s^{-1}}}

\newcommand{\msun}{\ensuremath{\rm M_{\odot}}}
\newcommand{\arcs}{\ensuremath{\rm arcsec}}
\newcommand{\cc}{\ensuremath{\rm cm^{-3}}}
\newcommand{\msunyr}{\ensuremath{\rm M_{\odot}\ yr^{-1}}}
\newcommand{\kelvin}{\ensuremath{\rm K}}
\newcommand{\yr}{\ensuremath{\rm yr}}

\newcommand{\erg}{\ensuremath{\rm erg}}
\newcommand{\rsun}{\ensuremath{\rm R_{\odot}}}

\usepackage{color}

\begin{document}

   \title{The formation and stability of a cold disc made out of stellar winds in the Galactic centre}
%   \subtitle{}
    \titlerunning{The formation and stability of a cold in the Galactic Centre}
    \authorrunning{Calderón et al.}

   \author{
        Diego Calderón\inst{\ref{inst1}}\fnmsep\inst{\ref{inst2}}\fnmsep\thanks{Alexander von Humboldt Fellow}\orcidlink{0000-0002-9019-9951}
        \and
        Jorge Cuadra\inst{\ref{inst3}}\fnmsep\inst{\ref{inst4}}\orcidlink{0000-0003-1965-3346}
        \and
        Christopher M. P. Russell\inst{\ref{inst5}}\orcidlink{0000-0002-9213-0763}
        \and
        Andreas Burkert\inst{\ref{inst6}}\fnmsep\inst{\ref{inst7}}\fnmsep\inst{\ref{inst8}}\orcidlink{0000-0001-6879-9822}
        \and
        \\
        Stephan Rosswog\inst{\ref{inst1}}\fnmsep\inst{\ref{inst9}}\orcidlink{0000-0002-3833-8520}
        \and
        Mayura Balakrishnan\inst{\ref{inst10}}\orcidlink{0000-0001-9641-6550} 
    }

   \institute{
        Hamburger Sternwarte, Universit\"at Hamburg, Gojenbergsweg 112, 21029 Hamburg, Germany\label{inst1}
        \and
        Max-Planck-Institut für Astrophysik, Karl-Schwarzschild-Straße 1, 85748 Garching, Germany\label{inst2}\\
        \email{calderon@mpa-garching.mpg.de}
        \and
        Departamento de Ciencias, Facultad de Artes Liberales, Universidad Adolfo Ib\'a\~nez, Av.\ Padre Hurtado 750, Vi\~na del Mar, Chile\label{inst3}
        \and
        Millennium Nucleus on Transversal Research and Technology to Explore Supermassive Black Holes (TITANS), Chile\label{inst4}
        \and
        Department of Physics and Astronomy, Bartol Research Institute, University of Delaware, Newark, DE 19716, USA\label{inst5}
        \and
        Universit\"ats-Sternwarte, Ludwig-Maximilians-Universit\"at M\"unchen, Scheinerstr. 1, 81679 Munich, Germany\label{inst6}
        \and
        Max-Planck Institute for Extraterrestrial Physics, Giessenbacherstr. 1, 85748 Garching, Germany\label{inst7}
        \and
        Excellence Cluster ORIGINS, Boltzmannstrasse 2, 85748 Garching, Germany\label{inst8}
        \and
        The Oskar Klein Centre, Department of Astronomy, AlbaNova, Stockholm University, SE-106 91 Stockholm, Sweden\label{inst9}
        \and
        Department of Astronomy, University of Michigan, 1085 S. University, Ann Arbor, MI 48109, USA\label{inst10}
    }

   \date{Received \today; accepted \today}

% \abstract{}{}{}{}{} 
% 5 {} token are mandatory
 
  \abstract
   {The reported discovery of a cold ($\sim$10$^4~\text{K}$) disc-like structure within the central $5\times10^{-3}$~pc around the super-massive black hole at the centre of the Milky Way, Sagittarius A* (Sgr~A*), has challenged our understanding of the gas dynamics and thermodynamic state of the plasma in its immediate vicinity. 
   State-of-the-art simulations do not agree on whether or not such a disc can indeed be a product of the multiple stellar wind interactions of the mass-losing stars in the region.}
   {The aims of this study are to constrain the conditions for the formation of a cold disc as a natural outcome of the system of the mass-losing stars orbiting around Sgr~A*, to investigate whether the disc is a transient or long-lasting structure, and to assess the validity of the model through direct comparisons with observations.}
   {We performed a set of hydrodynamic simulations of the observed Wolf-Rayet (WR) stars feeding Sgr~A* using the finite-volume adaptive mesh refinement code Ramses. 
   We focus, for the first time, on the impact of the chemical composition of the plasma emanating from the WR stars. 
   }
   {The simulations show that the chemical composition of the plasma affects the radiative cooling to a sufficient degree to impact the properties of the medium, such as density and temperature, and, as a consequence, the rate at which the material inflows onto Sgr~A*.
   We demonstrate that the formation of a cold disc from the stellar winds is possible for certain chemical compositions that are consistent with the current observational constraints. 
   However, even in such cases, it is not possible to reproduce the reported properties of the observed disc-like structure, namely its inclination and the  fluxes of its hydrogen recombination lines.} 
   {We conclude that the stellar winds alone are not sufficient to form the cold disc around Sgr~A* inferred from observations.
   Either relevant ingredients are still missing in the model, or the interpretation of the observed data needs to be revised.}
   \maketitle
%
%-------------------------------------------------------------------

\section{Introduction}

    The Galactic centre (GC) hosts the closest super-massive black hole to us, Sagittarius~A* \citep[Sgr~A*; see][for a review]{genzel2010}.  
    Unlike the black holes present in active galactic nuclei, Sgr~A* is highly underluminous.
    It does not seem to be currently accreting a significant amount of material or to have a standard accretion disc around it \citep{yuannarayan2014}. Nevertheless,
    \cite{murchikova2019} detected a disc-like structure around Sgr~A*.  
    Using the 1.3 millimetre recombination line H30$\alpha,$ these authors observed a double-peaked emission line with a full velocity width of $\sim$2200~\kms. 
    The centre of the emission coincides with Sgr~A* and extends up to 0.11\arcsec~(or $\sim10^4$ Schwarzschild radii) to both the redshifted and blueshifted sides. 
    \cite{murchikova2019} interpreted this feature as a rotating disc with a mass of $10^{-5}$-$10^{-4}$~\msun.
    \cite{yusef-zadeh2020} also reported the presence of broad hydrogen recombination lines ---including the double-peaked H$30\alpha$ line--- at the position of Sgr~A* using the Very Large Array. 
    However, these authors interpreted the signatures as a jet emanating from the central region. 
    So far, there has been no detection in the near-infrared, despite many observations. 
    \cite{ciurlo2021} report an upper limit for Br$\gamma$ that is two orders of magnitude below the extrapolation from the reported H$30\alpha$ flux. 
    At present, there is no consensus on how such observations can be reconciled or from where the observed cold material might come.

    The gaseous environment around Sgr~A* is dominated by the outflows from around 30 Wolf-Rayet (WR) stars, all located within a fraction of a parsec of the black hole \citep{paumard2006,martins2007}.  
    At such close distances, the winds interact strongly in shocks that thermalise them and create a hot and diffuse X-ray-emitting plasma \citep{quataert2004}. 
    However, given the relatively low velocity of some of the outflows \citep[450-600~km~s$^{-1}$;][]{martins2007}, the resulting plasma can be prone to radiative cooling and form dense, cold clumps \citep{cuadra2005,calderon2016}, which end up being embedded in the diffuse, hot plasma \citep[$\sim$10$^7$~K;][]{baganoff2003}. 
    The complex interplay between the stellar winds around Sgr A* has been the subject of hydrodynamic simulations using a variety of numerical techniques.
    With smoothed particle hydrodynamics (SPH) simulations, \cite{cuadra2006} showed that, if most of the slow-wind stars are located in a relatively compact stellar disc, many cold clumps form and quickly coalesce in a cold gaseous disc with a radius of $\sim$1\arcsec.  However, using a stellar distribution closer to the one observed, \cite{cuadra2008, cuadra2015} showed that no conspicuous disc appears over the $1800\,$yr time-span of their models. 
    \cite{calderon2020} performed finite-volume hydrodynamic simulations on a Cartesian grid of the same system ---which are better suited to modelling shocks, the multi-phase medium, and subsequent cooling--- and found that the formation of a cold disc is indeed possible. 
    According to their model, the stellar wind of the star IRS~33E, which is relatively dense ($\sim$10$^{-5}~\msunyr$) and slow \citep[$450~\kms$;][]{martins2007}, interacts with the medium, creating a shell that is dense enough to
quickly radiate its thermal energy and break into denser and smaller structures. 
    These pieces manage to fall inwards, forming a cold ($\sim$10$^4~\text{K}$) disc with a total mass of $\sim$5$\times10^{-3}~\msun$ within a simulation time of $3500~\text{yr}$. 
    Based on this result, \cite{ciurlo2021} found that the properties of the modelled disc are consistent with the non-detection in Br$\gamma$.
    Nevertheless, this scenario has not been confirmed by analogous grid-based simulations. 
    \cite{ressler2020} revisited the same system using the same numerical approach, although with a different code: \textsc{athena++} \citep{stone2020}, and found no disc formation, even when extending their simulation time up to $9000~\text{yr}$. 
    Building on this model, \cite{solanki2023} also studied this system through hydrodynamic modelling, but encompassing a larger region and evolving it for much longer timescales.
    These authors included the presence of the circumnuclear disc (CND), an observed gaseous structure located between 1.5 and $3.0$~pc~\citep{becklin1982,genzel1989} and with a total mass of $3$-$4\times10^4$~\msun~\citep{dinh2021,james2021}. 
    As a result, they showed that the formation of a cold disc is possible on long timescales ($\gtrsim$300,000~\yr) due to the interaction of the innermost boundary of the CND and the WR stellar winds, which results in the transport of CND material to smaller scales. 
    However, the authors warned that the lack of certain physical mechanisms in the model on such timescales (e.g. magnetic fields, supernovae, and thermal conduction) might affect the robustness of this result. 
    Thus, the exact formation process of the observed cold disc remains unknown.

    A key factor driving the formation of clumps and a disc in the model of \cite{calderon2020} is that radiative cooling allows the gas to get rid of its energy efficiently.  
    The strength of this process depends on the chemical composition of the gas, which is highly unusual in the GC given its origin as winds from evolved massive stars.  
    This work explores, for the first time, the impact of radiative cooling through studying specific chemical compositions: solar with $1~Z_{\odot}$, $3~Z_{\odot}$, and $5~Z_{\odot}$.
    More importantly, we developed a more realistic chemical setup that considers the atmospheric abundances for the different WR subtypes present in the region in order to follow the thermodynamic evolution of the gas in a more appropriate manner.
    Our results show that the formation of the disc is indeed determined by the composition of the gas. 
    However, the properties of this disc do not agree with the observations when considering realistic wind compositions compatible with current observational constraints.
    We also improved the numerical setup in order to better assess the disc orientation and more directly compare our results with those of \cite{ressler2018,ressler2020} as well as with SPH models \citep{cuadra2008,cuadra2015,russell2017}.
     
    This article is organised as follows: 
    In Section~\ref{sec:sims}, we present the numerical approach, the setup, and the models investigated. 
    Section~\ref{sec:results} presents and describes the results of the numerical simulations. 
    In Section~\ref{sec:mock}, we present the synthetic observables obtained through post-processing the models and contrast them with observations. 
    We compare our grid-based models with SPH models in Section~\ref{sec:sph}. 
    In Section~\ref{sec:stability}, we present an analytic analysis of the stability of the observed cold disc.
    In Section~\ref{sec:discussion}, we discuss our findings and uncertainties in the parameters of our models. 
    Finally, in Section~\ref{sec:conclusions}, we present conclusions and final remarks.
    Throughout this paper we use the mass and distance to Sgr~A* of $4.3\times10^6~\msun$ and $8.33~\text{kpc}$, respectively \citep{gillessen2017,gravity2019,gravity2021}, so that $1\arcsec$ corresponds to a length of $\sim$0.04~pc~$\approx10^5$~R$_\text{Sch}$, where R$_\text{Sch}$ is the Schwarzschild radius.

%--------------------------------------------------------------------
\section{Numerical simulations}
\label{sec:sims}
    \subsection{Equations}
        The simulations were performed using the adaptive-mesh refinement (AMR) hydrodynamic code Ramses \citep{teyssier2002}. 
         This code uses a second-order Godunov method with a shock-capturing scheme to solve the Euler equations in their conservative form, that is,
        \begin{eqnarray}
            \frac{\partial\rho}{\partial t}+\nabla\cdot\left(\rho \mathbf{u}\right)
            &=&
            0,
            \label{eq:mass}
            \\
            \frac{\partial}{\partial t}\left(\rho\mathbf{u}\right)+\nabla\cdot\left(\rho\mathbf{u}\mathbf{u}\right)
            &=&
            -\nabla p-\rho\nabla\phi,
            \label{eq:momentum}
            \\
            \frac{\partial}{\partial t}\left(\rho e\right)+\nabla\cdot\left[\rho\mathbf{u}\left(e+\frac{p}{\rho}\right)\right]
            &=&
            -\rho\mathbf{u}\cdot\nabla\phi-Q^{-}_\text{tot},
            \label{eq:energy}
            \\
            \frac{\partial}{\partial t}\left(\rho s_i\right)+\nabla\cdot\left(\rho s_i\mathbf{u}\right)
            &=&
            0
            \label{eq:tracer}
        ,\end{eqnarray}
        where $(\rho,\mathbf{u},P)$ are the primitive hydrodynamic variables: density, velocity, and pressure, respectively. 
        The set of quantities $s_i$ correspond to tracer scalar fields that are advected with the fluid whose usage we introduce in Section~\ref{sec:models}.
        The sink terms on the right-hand side correspond to the effects of the time-independent gravitational potential $\phi=\phi(\mathbf{x})$ (in our case solely of Sgr~A*; see Section~\ref{sec:setup}), and the total radiative losses due to optically thin radiative cooling $Q^{-}_\text{tot}=Q^-_\text{tot}(\rho,T,X_\text{i})$, with $\mathbf{x}$ the position vector, $T$ the fluid temperature, and $X_\text{i}$ the chemical composition of the fluid. 
        The total specific energy density $e$ is given by
        \begin{equation}
            e=\frac{1}{2}\mathbf{u}\cdot\mathbf{u}+\frac{p}{(\gamma-1)\rho},
        \end{equation}
        where $\gamma$ is the adiabatic index that is set to $5/3$. 
        Additionally, we consider the fluid can be described as an ideal gas so that the temperature can be calculated through $P=(\rho/\mu)k_\text{B}T$, being $\mu$ the mean molecular weight and $k_\text{B}$ the Boltzmann constant. 

    \begin{table*}
    \begin{center}
    \begin{threeparttable}
        \caption{Atmospheric mass fractions (in percentage) and mean molecular weights.}
        \begin{tabular}{lccccccc}
            \hline
            \\
            Composition
            &$X_\text{H}$ & $X_\text{He}$ & $X_\text{C}$ & $X_\text{N}$ & $X_\text{O}$ & $\sum_{\text{i}\neq\text{H,He,C,N,O}} X_\text{i}$ & $\mu_{\rm ion}$
             \\
            (1)&(2)&(3)&(4)&(5)&(6)&(7)&(8)
            \\
            \hline
            \\
            Solar              & 74.91 & 23.77  & 0.2191 & 0.0704 & 0.5824 & 0.4481 & 0.5942
            \\
            WC8-9              & 0     & 60     & 31     & 0      & 7      & 2 & 1.5385
            \\
            WN5-7              & 0     & 98.5   & 0.029  & 1      & 0.018  & 0.453 & 1.3536
            \\
            WN8-9 and Ofpe/WN9 & 11.5  & 82.4   & 0.0124 & 1.15   & 0.066  & 4.872 & 1.1383
            \\
            \hline
        \end{tabular}
        \label{tab:abundances}
    \begin{tablenotes}
        \item
        \textit{Notes.} 
        The solar abundances were taken from \cite{lodders2003}. 
        The abundances for the three WR subtypes correspond to the compilation by \cite{russell2017} based on previous studies \citep{herald2001,crowther2007,onifer2008}. 
        Column~1: chemical mixture. 
        Columns~2-7: hydrogen, helium, carbon, nitrogen, oxygen, and rest of the metals mass fractions, respectively. 
        Column~8: mean molecular weigh assuming full ionisation.
    \end{tablenotes}
    \end{threeparttable}
    \end{center}
    \end{table*}

        \begin{table}
            \centering
            \begin{threeparttable}
            \caption{Simulation runs and parameters.}
            \begin{tabular}{lcc}
                \hline
                \\
                Simulation & Composition &   Disc\\
                (1) & (2) & (3)
                \\
                \hline
                A1       &  Solar                     &   No\\
                A3       &  $\text{Solar}+3Z_{\odot}$ &   Yes\\
                A5       &  $\text{Solar}+5Z_{\odot}$ &   Yes\\
                WR\_f07 &  Differential               &   No\\
                WR\_f1  &  $\text{Differential~with}~X_\text{H}=40$ &   Yes\\
                \hline
                \hline
            \end{tabular}
            \label{tab:sims}
            \begin{tablenotes}
                \item
                \textit{Notes.} 
                Column~1: simulation ID. 
                Column~2: chemical composition of the winds. 
                Column~3: whether or not the final state of the system shows the presence of a cold disc around Sgr~A*.
            \end{tablenotes}
            \end{threeparttable}
        \end{table}

    \subsection{Numerical setup}
    \label{sec:setup}

        The model considers the system of WR stars blowing stellar winds while they move on their observed Keplerian orbits around Sgr~A*. 
        The setup is analogous to our previous work \citep{calderon2020}. 
        The central black hole gravitational field is modelled as a point mass of $4.3\times10^6$~\msun~\citep{gravity2019,gravity2021}. 
        The stars are simulated as test particles that only feel the gravitational pull of Sgr~A*. 
        Their initial position and velocity vectors are set so that they move on the orbits constrained by observations \citep{paumard2006,cuadra2008,gillessen2019,vonFellenberg2022}. 
        The stellar winds in the simulation are modelled following the procedure by \cite{lemaster2007}, which has been validated and used in our previous work \citep{calderon2020b,calderon2020}. 
        This consists in defining a ``masked region" around each star that is a spherical volume where the hydrodynamic variables are reset in each timestep, in order to reproduce the free wind expansion solution of a spherical wind with certain mass-loss rate $\dot{M}_\text{w}$, terminal velocity $V_\text{w}$, and temperature $T_\text{w}$. 
        For these quantities we used the values  constrained through modelling of the infrared spectra \citep{martins2007,cuadra2008}. 
        The wind temperature was set to the lowest temperature allowed in the simulation, $T_\text{w}=10^4~\text{K}$
        which is determined by the strong ultraviolet radiation produced by the hundreds of massive stars in the region.
        
        The simulations were run in a Cartesian grid in a cubic domain of side 40\arcsec~($\sim$1.6~pc) with outflow boundary conditions (zero gradients). 
        We used the exact Riemann solver \citep[e.g.][]{toro2009} combined with the MinMod slope limiter, as these choices allow the modelling of hydrodynamic instabilities, which otherwise may be quenched due to numerical diffusion. 
        We investigated other choices such as the Harten-Lax-van Leer-Contact \citep[HLLC;][]{toro1994} and/or the MonCen slope limiter. 
        Our tests showed that the use of the HLLC solver amplified the grid artifacts while the use of the MonCen limiter resulted in unwanted numerical artifacts such as spurious oscillations.
        For a careful analysis of these choices we refer the reader to the Appendix~\ref{app:solver}. 
        Regarding the numerical resolution of our models, the coarse resolution was $64^3$ cells, allowing adaptive refinement of four levels, that is, effectively $\Delta x\approx0.0391\arcsec$ $\approx1.56\times10^{-3}$~pc. 
        However, the regions around the stars allowed two extra levels of refinement ($\Delta x\approx9.77~\text{mas}\approx3.91\times10^{}$~pc) in order to guarantee at least eight cells as the radius of the masked regions where the winds are initialised. 
        These regions have a fixed radius of $80~\text{mas}\approx3.2\times10^{-3}$~pc.
        Additionally, the vicinity around the central black hole has a fixed nested grid with eight refinement levels above the coarse resolution ($\Delta x\approx2.44~\text{mas}\approx9.77\times10^{-5}$~pc). 
        At the location of Sgr~A*, we defined a spherical region where the hydrodynamic variables are reset to low values of density and pressure at rest, in order to avoid artificial accumulation of material. 
        The refinement strategy is based on density gradients on top of the geometric criteria previously defined. 
        In order to explore the role of the AMR potentially affecting the results we tested different values of the smoothing parameter that controls the refinement in transitions between refinement levels. 
        We tested quadrupling the smoothing parameter and the results remained unchanged\footnote{The parameter is \texttt{nexpand} and controls the number of cells forced to be refined despite not being flagged for refinement but for being neighbouring cells of a refined cell. For instance, we tested \texttt{nexpand}=1,4 across all transitions}.

    \subsection{Models}
        \label{sec:models}
        The main parameter explored in this work is the optically thin radiative cooling function. 
        This choice is determined by specifying the chemical composition of the fluid.  
        Unfortunately, the metallicity of the young stars in the GC is still poorly known, yet it has been argued that it should be higher than Solar \citep[$Z=2-3~Z_{\odot}$;][]{genzel2010}. 
        Past numerical works have considered that the composition of the gas correspond to Solar abundances but with metallicity three times the Solar value, $Z=3~Z_\odot$ \citep{cuadra2008,calderon2016,ressler2018,ressler2020,solanki2023}. 
        However, more than the `bulk' metallicity of the stars, the most appropriate composition choice would be the abundances of the WR stellar atmospheres, which normally differ significantly from Solar \citep{herald2001,onifer2008}. 
        Although \cite{ressler2018,ressler2020} used chemical compositions lacking hydrogen, the rest of the element abundances remained unchanged relative to Solar with $3Z_{\odot}$. 
        As a result, the cooling function varied at low temperature (<10$^5~\text{K}$) but at higher temperatures it remains unchanged. 
        Moreover, the composition choice not only determines the cooling function but also the ion mean molecular weight as well as the electron to proton number densities $n_\text{e}/n_\text{p}$. 
        These quantities are taken into account in the sink term in the energy equation (see equation~\ref{eq:energy}), and following \cite{schure2009} can be expressed as follows
        \begin{equation}
            Q^{-}_i(\rho,T_i,X_i)=n_\text{p}^2\left(\frac{n_\text{e}}{n_\text{p}}\right)_i\Lambda_i(T_i),
        \end{equation}
        where the subscript $i$ stands for a given chemical abundance. 
        Bear in mind that different chemical compositions also correspond to different temperature $T_i$, as this is obtained through the ideal gas expression that makes uses of the mean molecular weight of the fluid. 
        In the general case, if we consider a fluid element composed of $N$ mixtures of abundances, the total radiative losses will be given as a summation, that is,
        \begin{equation}
            Q^{-}_\text{tot}(\rho,T_i,X_i)=n_\text{p}^2\frac{\sum_{i=1}^Ns_i\left(\frac{n_\text{e}}{n_\text{p}}\right)_i\Lambda_i(T_i)}{\sum^N_{i=1}s_i},
        \end{equation}
        where we have expressed the total radative cooling as a linear combination of the mixtures. 
        Notice that we have introduced the passive scalar fields $s_i$, as we used them to quantify the fraction of a given chemical composition. 
        In this work, we introduced three types of compositions that represent the WR subtypes based on their atmosphere abundances that we assume to be constant over the few thousand year duration of the simulations. 
        In the region where the winds are generated, the corresponding passive scalar is set to $1$ while the rest are set to $0$. 
        By doing so, it is possible to identify how much material of a given cell is supplied by which subgroup of WR stars. 
        The mass fractions of each mixture are shown in Table~\ref{tab:abundances}. 
        
\begin{figure}
        \centering
        \includegraphics[width=0.52\textwidth]{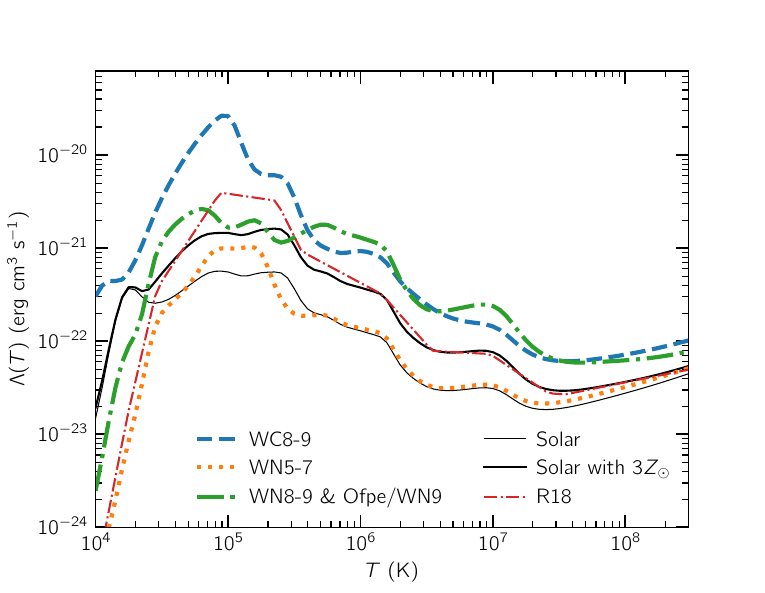}
        \caption{
        Comparison of cooling functions $\Lambda(T)$ for different chemical abundances. 
        The thin and thick lines show the radiative cooling for Solar and three times Solar abundances, respectively. 
        The dashed blue, dotted orange, and dot-dashed green lines represent the cooling functions for atmosphere abundances corresponding to WR subtypes: WC8-9, WN5-7,  WN8-9 and Ofpe/WN9 based on the compositions compiled shown in Table~\ref{tab:abundances}.
        The contributions of each element to the radiative cooling were taken from the plasma models by \cite{stofanova2021}. 
        Additionally, the cooling function used in \cite{ressler2018} is shown as a thick black line.
        }
        \label{fig:lambdas}
    \end{figure}

        In this work, we investigated five models with different compositions: Solar composition with metallicities $Z_{\odot}$, $3Z_{\odot}$, and $5Z_{\odot}$, a mixture of three compositions based on the spectroscopic constraints of the subtypes of WR stars, and a variation of the latter one motivated by the uncertainty in the WR stellar atmospheres and, specifically on the H fraction in them that is set to $X_\text{H}=40\%$ (see Section~\ref{sec:H}, for a discussion). 
        Figure~\ref{fig:lambdas} shows the cooling function $\Lambda=\Lambda(T)$ as a function of temperature for different chemical mixtures. 
        Each curve was calculated as a linear combination of the abundance of a given element and its contribution to the total radiative cooling. 
        The values of the composition per element are shown in Table~\ref{tab:abundances}.
        The contribution of each element to the total cooling was taken from the plasma models developed by \cite{stofanova2021}. 
        The cooling functions corresponding to WR subtypes compositions: WC8-9, WN5-7, and WN8-9/Ofpe are shown with dashed blue, dotted orange, and dotted-dashed green lines, respectively. 
        The Solar and Solar with $3Z_{\odot}$ are represented with thin and thick solid black lines, respectively. 
        Furthermore, we added the cooling function used in previous works by \cite{ressler2018,ressler2020} as a dotted-dashed red line. 
        Here it can be observed that a Solar composition with $3Z_{\odot}$ (the typical used value) is between the range of the different WR compositions but the subtypes WC89 and WN89/Ofpe are about a factor two higher.  
        
        The list of the runs investigated in this work are shown in Table~\ref{tab:sims}. 
        The initial setup was identical to our previous work \citep{calderon2020}, i.e. the medium across the whole domain was set to constant and low-enough values of density $\rho=10^{-24}~\text{g}~\text{cm}^{-3}$ and pressure $P=\gamma^{-1}\rho c_\text{s,f}^2$ with $c_\text{s,f}=10~\text{km}~\text{s}^{-1}$, so that the stellar winds do not encounter impediments and fill quickly the domain.   
        All models were run for a total of 3,500~yr but setting the starting state of the system in the past, so that the final state of the simulations corresponds to the current state of the system. 
        This is achieved by integrating the current position and velocity vectors of the stars back in time, assuming that they have followed purely Keplerian orbits \citep{paumard2006,vonFellenberg2022} within this timescale due to the gravitational field of Sgr A* \citep{cuadra2008,ressler2018,calderon2020}. 

    \begin{figure*}
        \centering
        \includegraphics[width=0.6\textwidth]{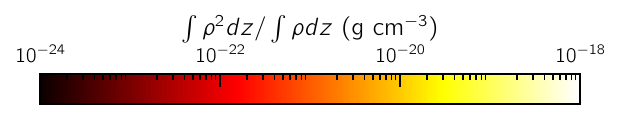}
        \includegraphics[width=0.475\textwidth]{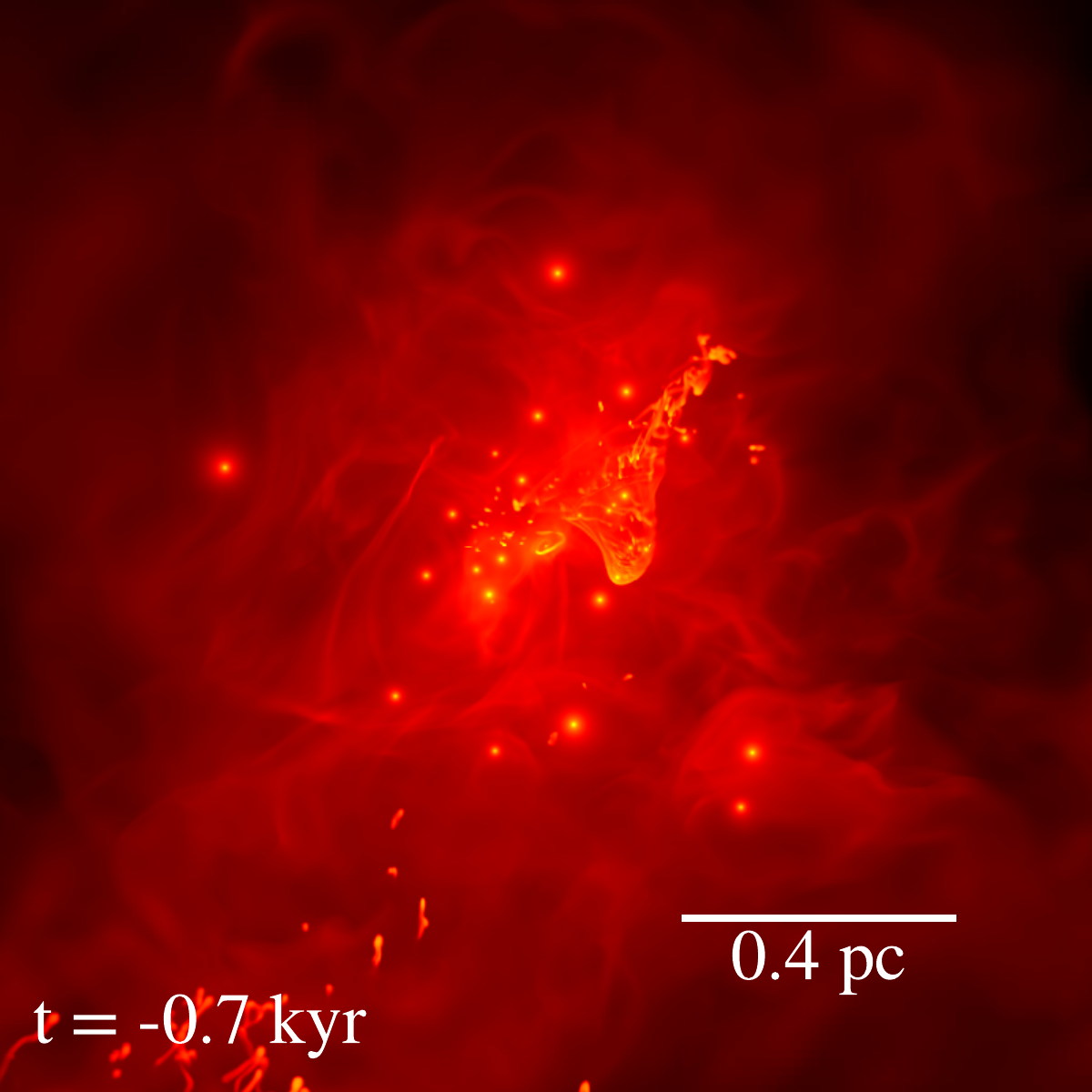}
        \includegraphics[width=0.475\textwidth]{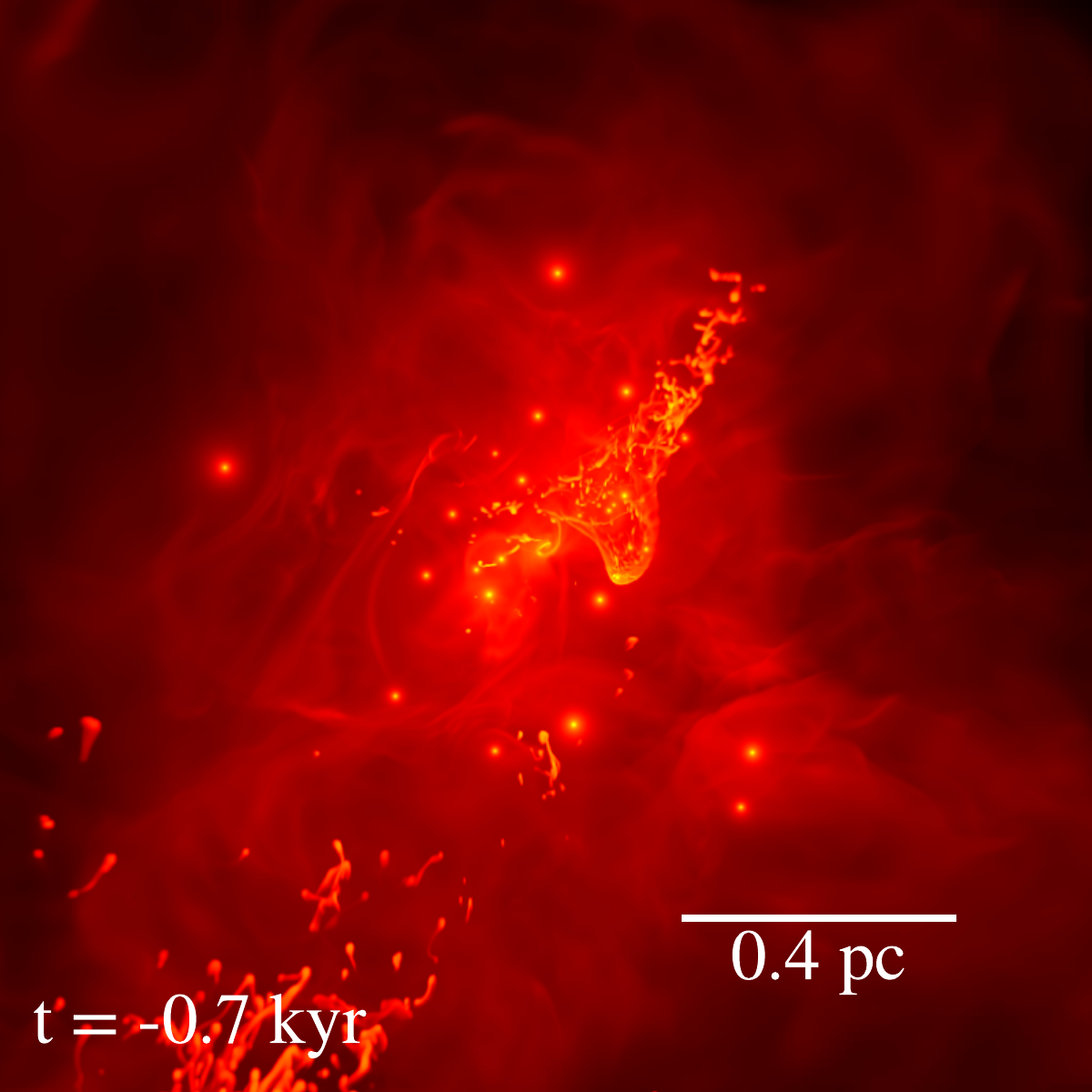}
        \includegraphics[width=0.475\textwidth]{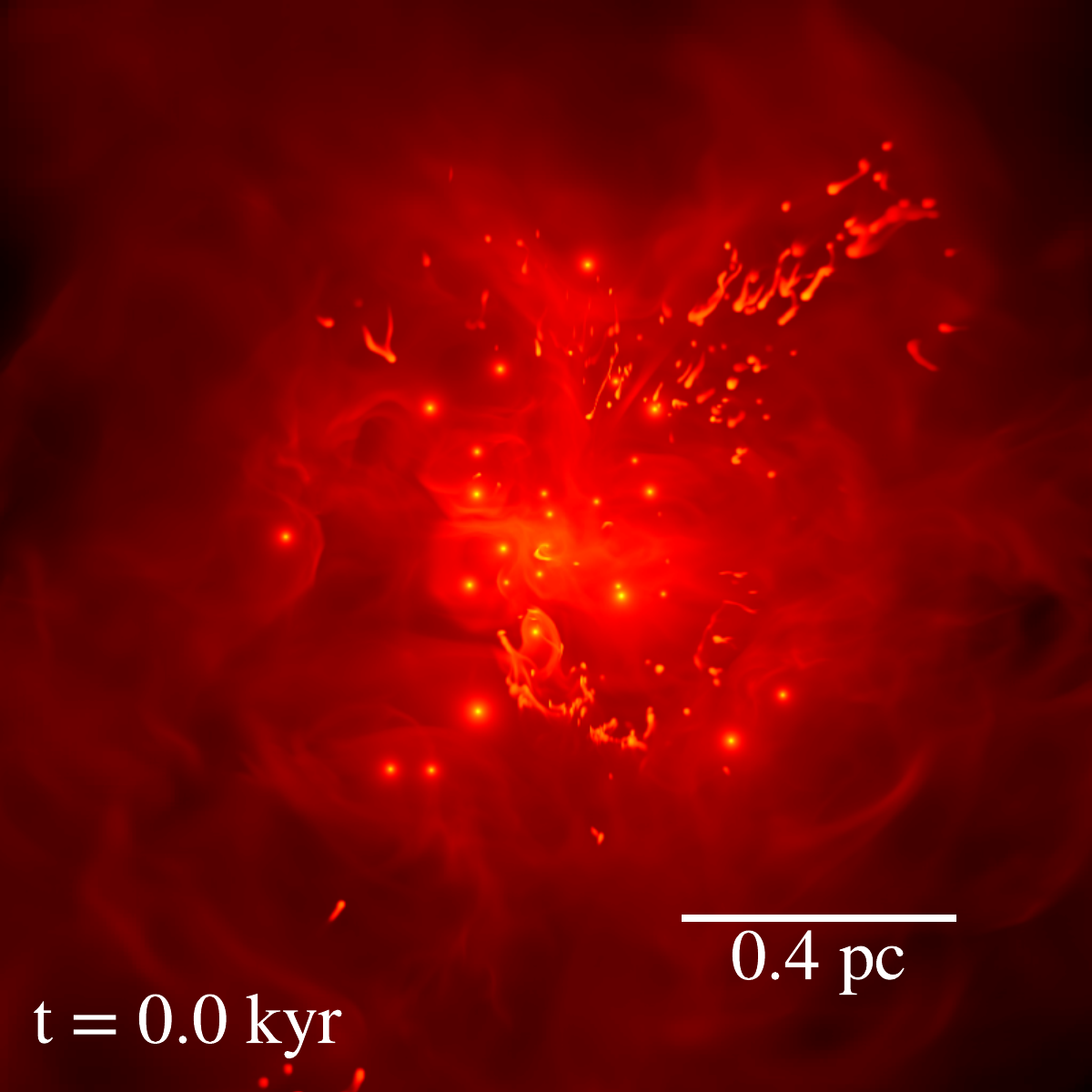}
        \includegraphics[width=0.475\textwidth]{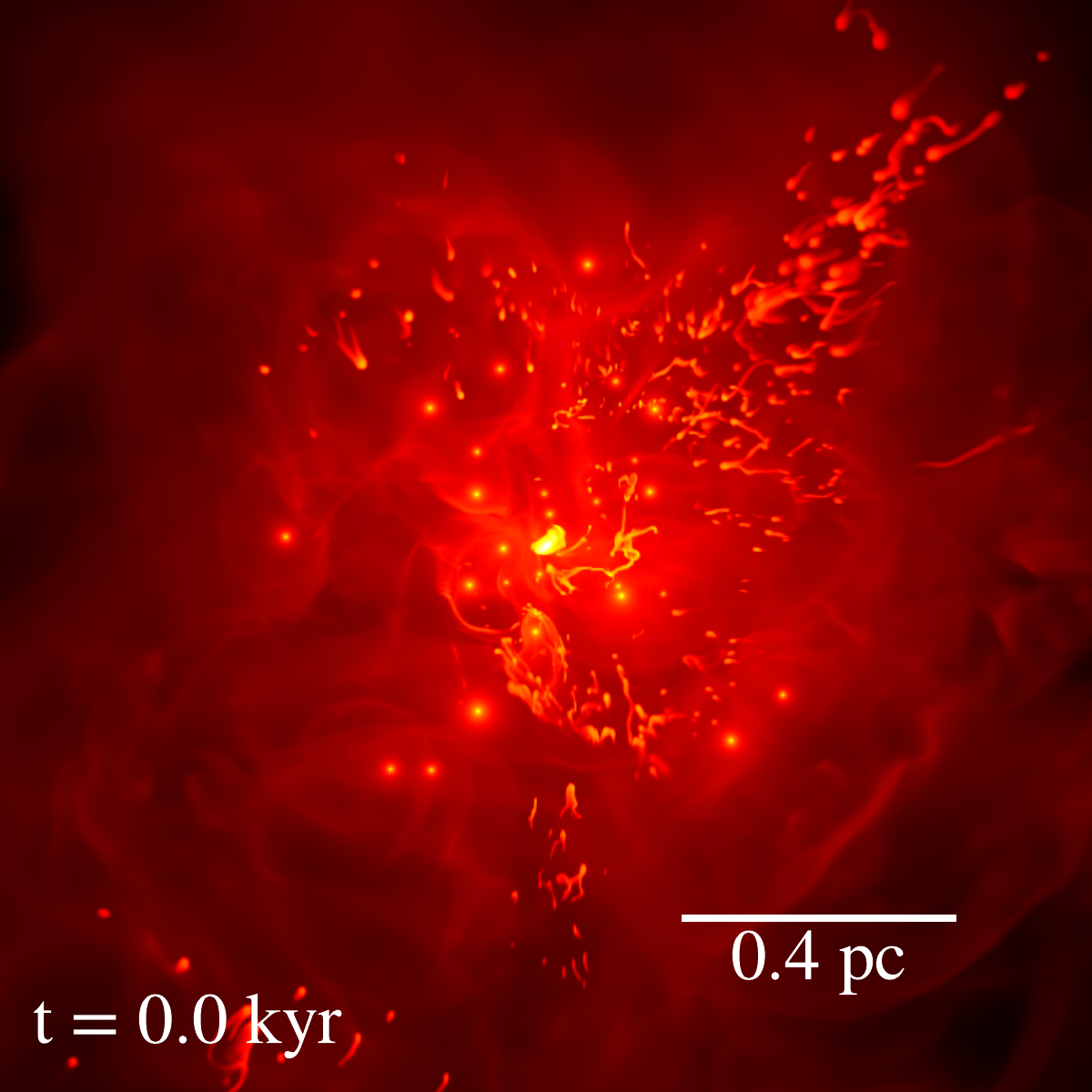}
        \caption{
        Comparison of the simulations with different chemical abundances at two different simulation times.  
        The panels show projected density maps weighed by density along the $z$ axis, i.e. $\int \rho^2dz/\int\rho dz$, which is parallel to the line of sight. 
        Top and bottom panels show the systems at $t=-0.7~\text{kyr}$ and $t=0$ (present), respectively. 
        Left- and right-hand side panels show the runs WR\_f07 and WR\_f1, respectively. 
        All maps display the full computational domain.}
        \label{fig:rho_maps}
    \end{figure*}

        \begin{figure}
            \centering
            \includegraphics[width=0.495\textwidth]{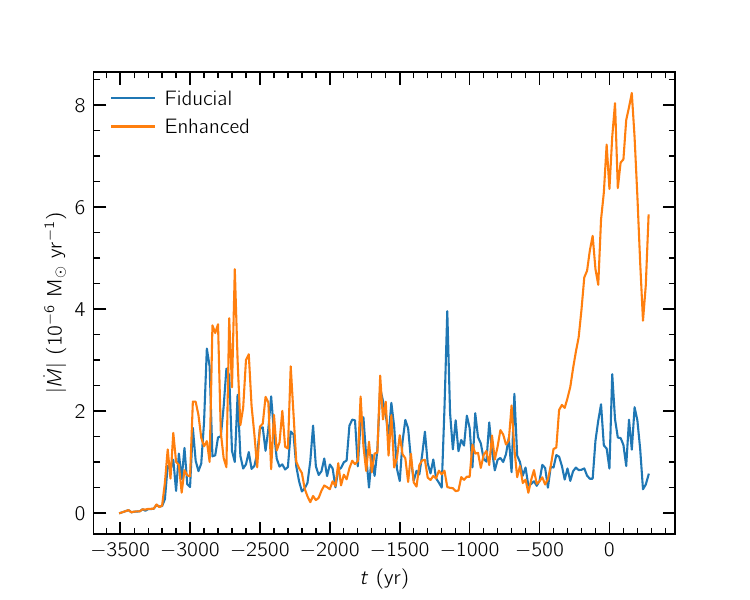}
            \caption{
            Net mass-flow rate across a sphere of radius $5r_\text{in}=5\times10^{-4}~\text{pc}$ ($1.25\times10^{-2}$\arcsec). 
            At this radius, the direction of the mass flow is inwards throughout the entire simulation. 
            The models WR\_f07 and WR\_f1 are shown in solid blue and orange lines, respectively.
            }
            \label{fig:mdot}
        \end{figure}

        \begin{figure}
            \centering
            \includegraphics[width=0.495\textwidth]{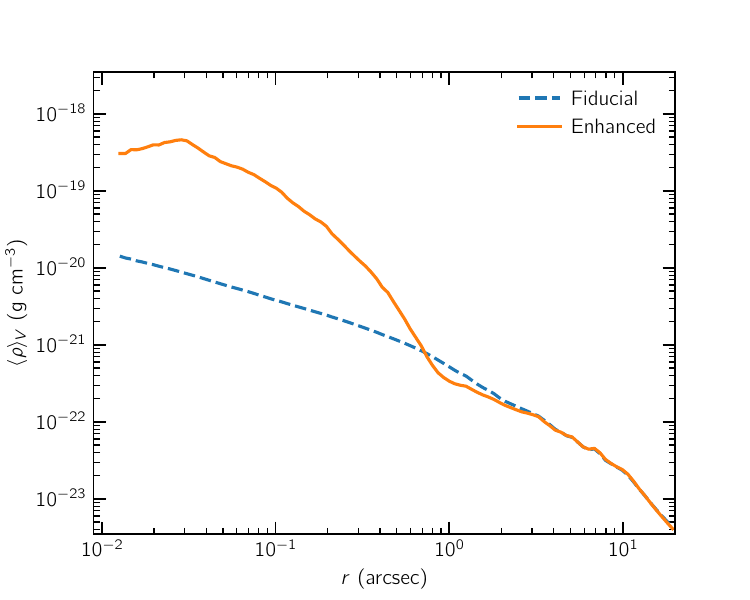}
            \includegraphics[width=0.495\textwidth]{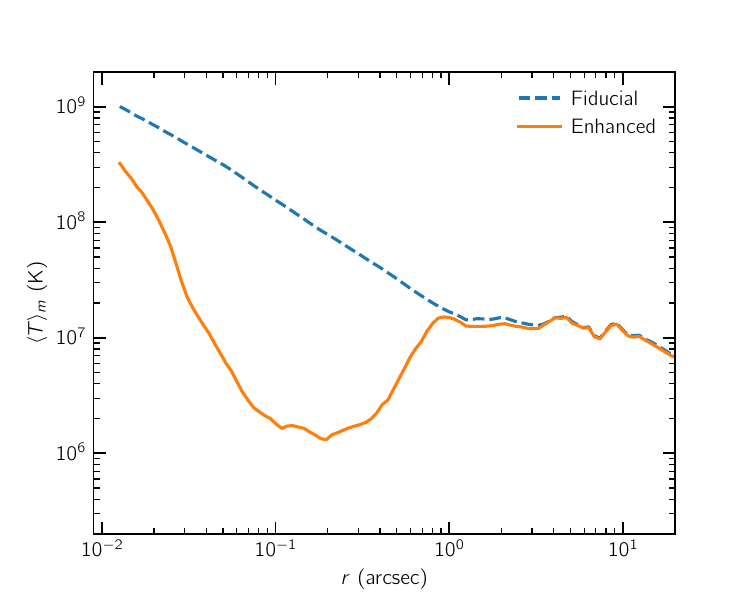}
            \caption{
            Radial profiles of time-averaged volume-weighted density (top) and mass-weighted temperature  (bottom) over the last 500~yr of simulation time. 
            The models WR\_f07 and WR\_f1 are shown in solid orange and dashed blue lines, respectively.
            }
            \label{fig:prof_rhoT}
        \end{figure}

        \begin{figure}
            \centering
            \includegraphics[width=0.495\textwidth]{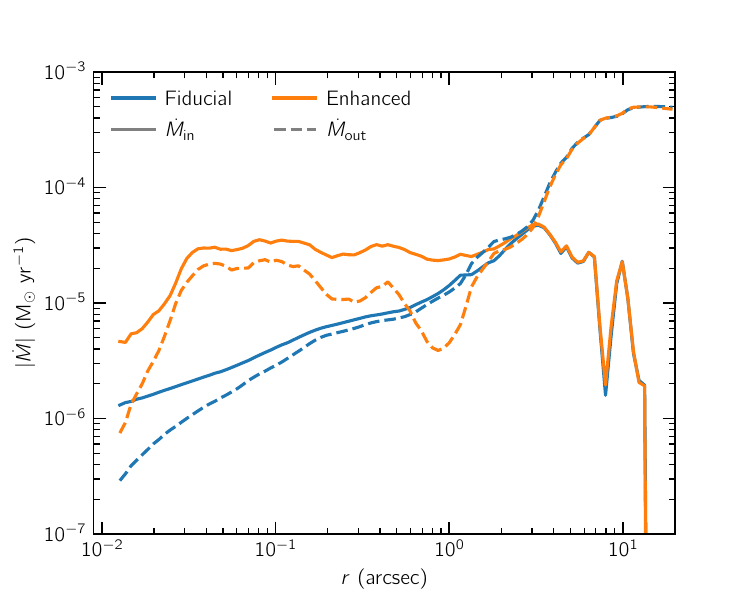}
            \caption{
            Radial profiles of the mass inflow $\dot{M}_\text{in}$ (solid lines) and outflow $\dot{M}_\text{out}$ (dashed lines) rates averaged over the last 500~yr of the simulation. 
            The models WR\_f07 and WR\_f1 are shown in solid orange and blue lines, respectively.
            }
            \label{fig:mdot_profile}
        \end{figure}

        \begin{figure}
            \centering
            \includegraphics[width=0.495\textwidth]{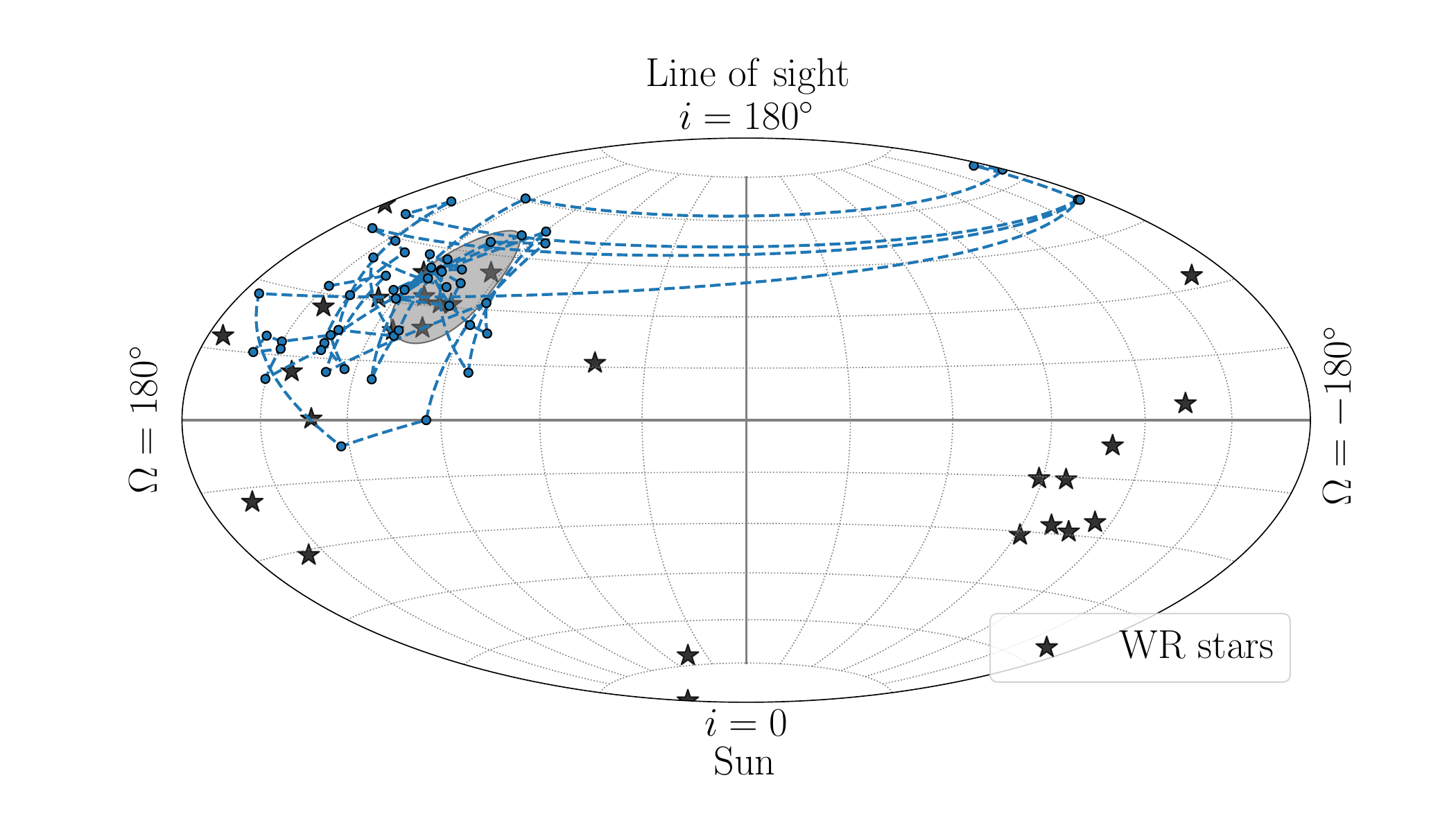}
            \includegraphics[width=0.495\textwidth]{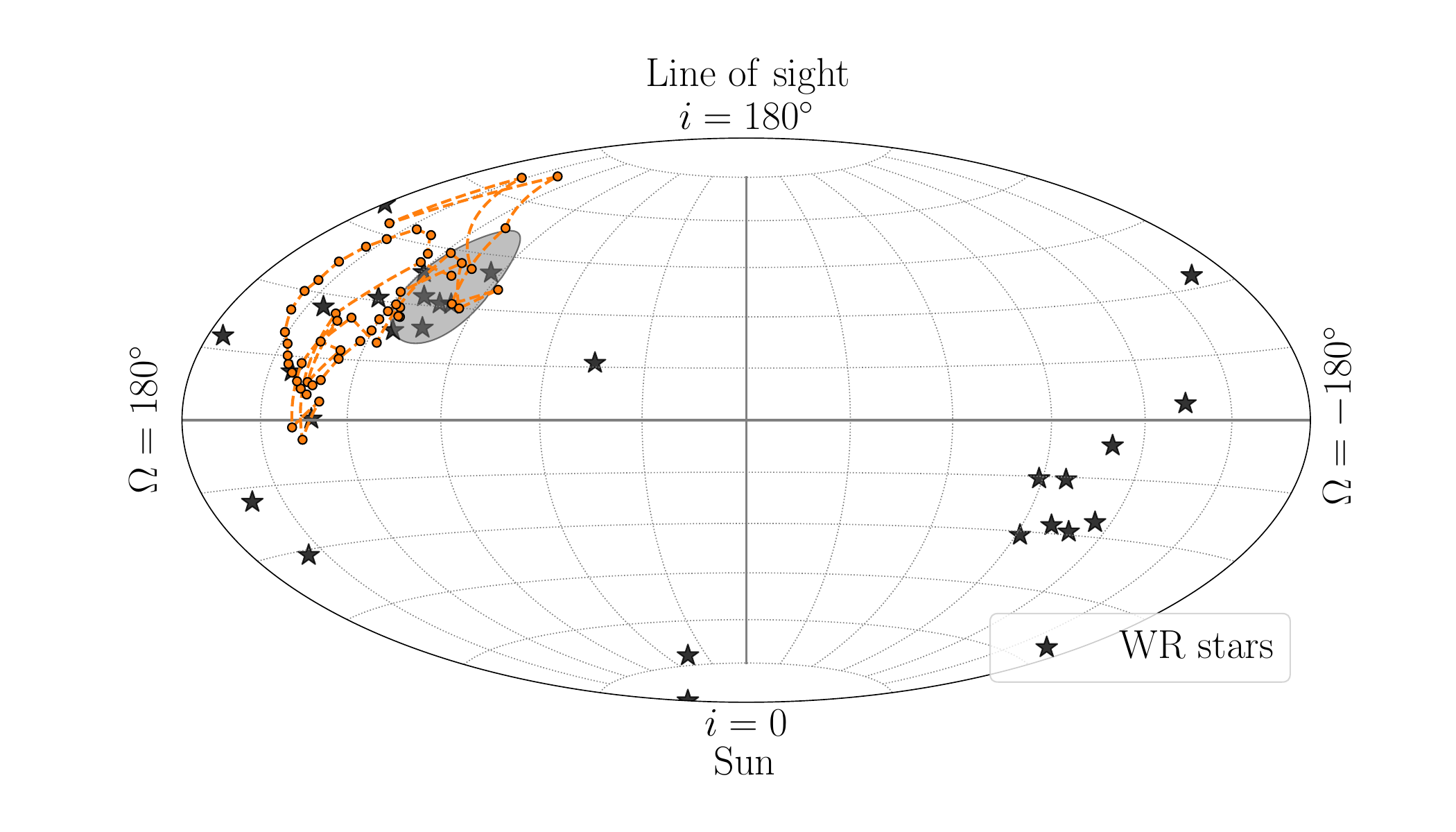}
            \caption{
            Orientation of the angular momentum of the gas enclosed in a sphere of radius 0.01~pc ($\sim$0.25\arcsec). 
            The vertical dimension represents inclination $i$, while the horizontal dimensions stand for the longitude of the ascending node $\Omega$. 
            Thus, a face-on star orbiting clockwise on the sky would be at the north pole of the graph. 
            Top and bottom panels show the runs WR\_f07 and WR\_f1, respectively.
            Each dot corresponds to the analysis of a single snapshot. 
            As a reference, the orientation of the orbital angular momentum of the mass-losing stars is shown as black star symbols, and the grey shaded region corresponds to the average direction of the clockwise disc at (104$^\circ$, 126$^{\circ}$) with a dispersion of 16$^\circ$ \citep{yelda2014}.
            }
            \label{fig:hammer}
        \end{figure}

        \begin{figure}
            \centering
            \includegraphics[width=0.495\textwidth]{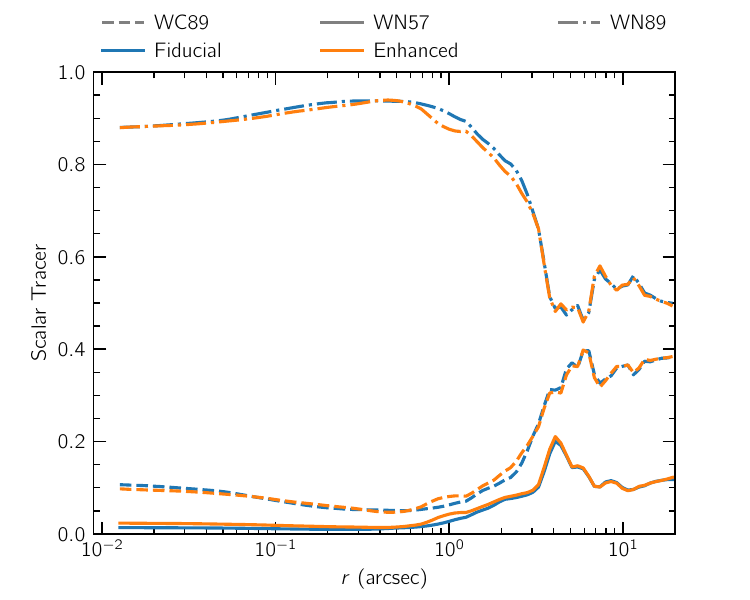}
            \caption{
            Radial profiles of the scalar tracer fields that represent the WR subtypes: WC89 (dashed lines), WN57 (solid lines), and WN89/Ofpe (dot-dashed) averaged over the last 500~yr of the simulation. 
            The models WR\_f07 and WR\_f1 are shown in solid orange and blue lines, respectively.
            }
            \label{fig:prof_tracer}
        \end{figure}
    
\section{Results}
\label{sec:results}
    We proceed to describe the evolution and final state of the simulations at $t=0$ that corresponds to the present time. 
    The runs with Solar composition, but varying metallicities, were used for understanding the impact of increasing and decreasing the effect of radiative cooling in general.
    However, since they do not represent realistic compositions we opted not to discuss them in detail here. 
    Instead, we chose to focus on the more physically motivated models, i.e. WR\_f07 and WR\_f1 which we also refer as \textit{fiducial} and \textit{enhanced}, respectively. 
    For completeness, we give a brief description of the models A1, A2, and A3 in Appendix~\ref{app:solar}.

        \begin{figure*}
            \centering
            \includegraphics[width=0.495\textwidth]{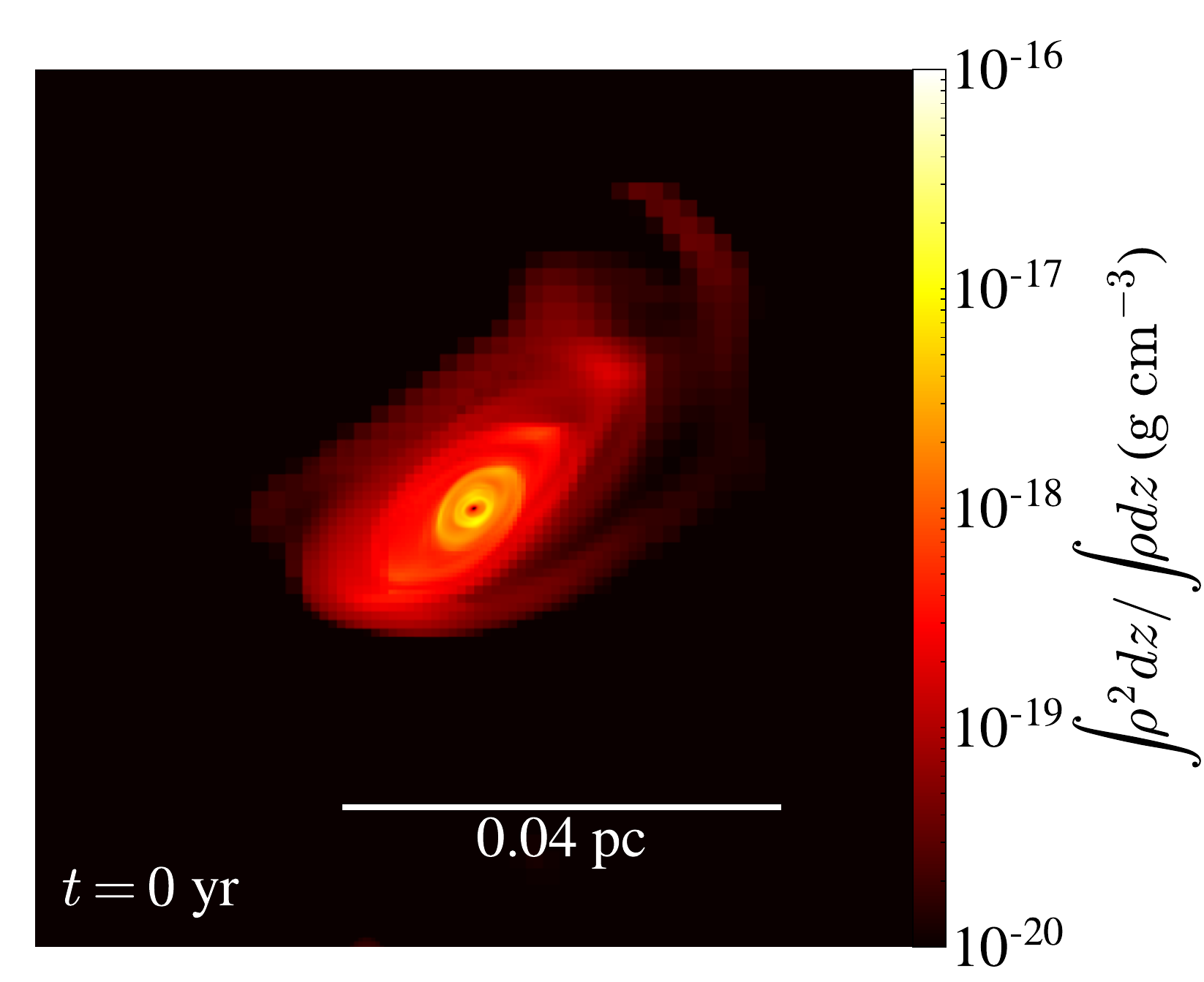}\includegraphics[width=0.495\textwidth]{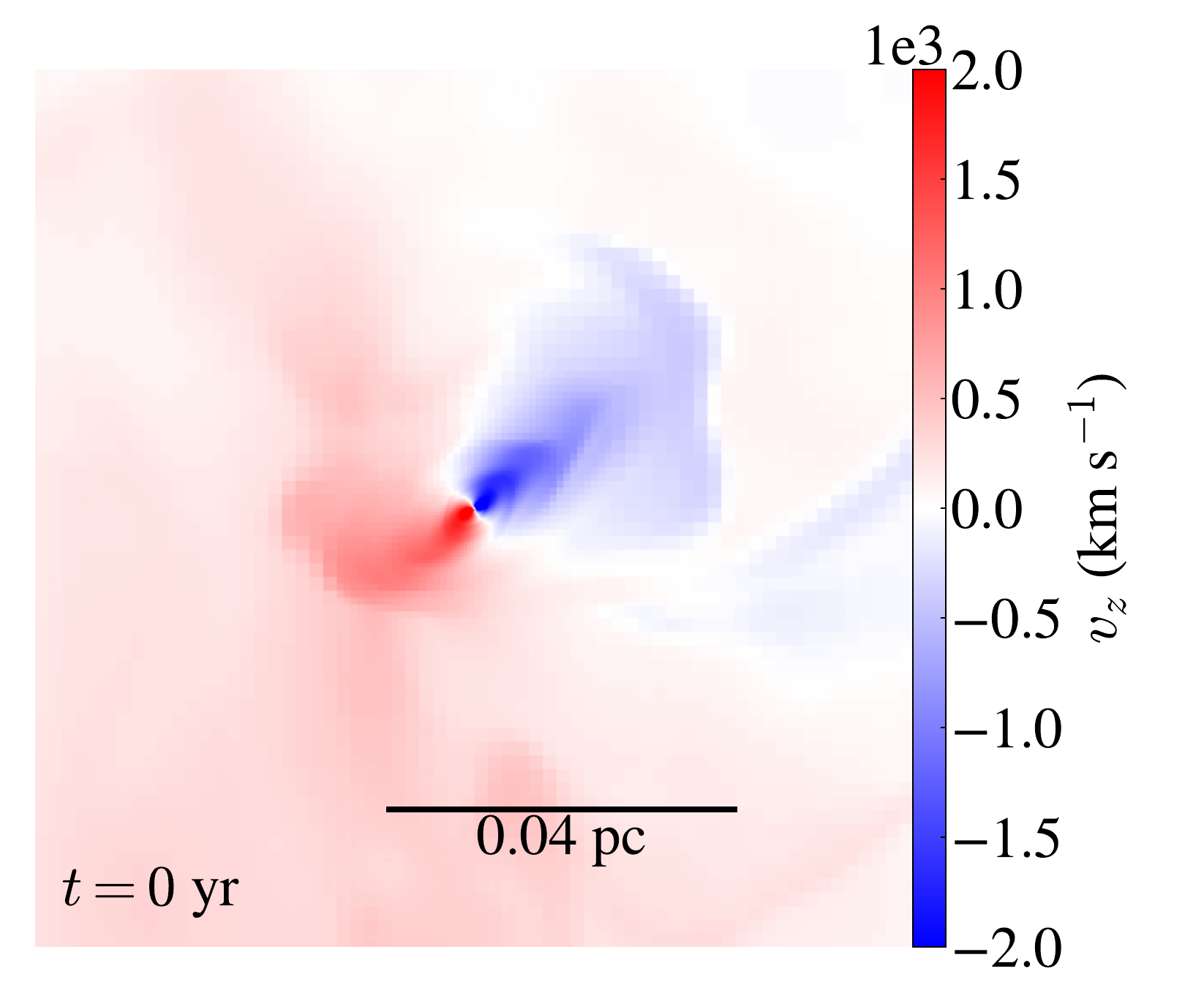}
            \caption{
            Density and line-of-sight velocity maps of the central 2.5\arcsec$\times$~2.5\arcsec of the model WR\_f1 at $t=0$. 
            The left- and right-hand side panels show projected density weighed by density and line-of-sight velocity weighed by mass, respectively both integrated along the z-axis, which is parallel to the line of sight.
            }
            \label{fig:disc_maps}
        \end{figure*}

    \subsection{Hydrodynamics}
    \label{sec:hydro}
        The simulation evolution of all models is analogous, especially in the initial phases. 
        At $t=-3500~\text{yr}$, the stars begin to orbit around Sgr~A* that is located at the centre of the domain, while their winds quickly fill the whole computational domain. 
        After $\sim500~\text{yr}$ ($t=-3000~\text{yr}$), the system reaches a quasi-steady state. 
        At this point, the domain is full of diffuse ($\sim10^{-22}-10^{-21}~\text{g}~\text{cm}^{-3}$) and hot plasma ($\sim10^7-10^8~\text{K}$) due to the shocked stellar winds and their interactions. 
        This is illustrated in Figure~\ref{fig:rho_maps} that shows density maps across different simulation times. 
        The maps show projected density fields along the line-of-sight direction ($z$ axis) weighted by density, i.e. $\int \rho^2dz/\int\rho dz$, in order to highlight the densest regions with the highest resolution. 
        The top panels show two models at $t=-700~\text{yr}$, respectively. 
        In them, it is possible to observe the complex structure developed due to the stellar wind interactions such as bow shocks, instabilities, and dense clumps as a result of condensation through radiative cooling. 
        Left- and right-hand side panels display models with different chemical compositions WR\_f07 and WR\_f1, respectively. 
        On the left-hand side, it can be seen how mildy efficient cooling affects mostly one of the stellar winds. 
        This star corresponds to IRS~33E and its wind is the slowest ($\sim$450$~\text{km}~\text{s}^{-1}$). 
        This fact causes its shocked temperature to be in an efficient region of the radiative cooling functions that allows its material to cool down and become denser. 
        Also, some of its material manages to reach the vicinity of Sgr~A*, as the map shows an elongated clump being accreted and stretched. 
        The right-hand side also portrays this picture but since the radiative cooling is enhanced more dense clumps can be observed, especially close to IRS~33E and Sgr~A*. 
        Finally, the bottom panels of Figure~\ref{fig:rho_maps} show the models at time $t=0$, i.e. the present time.  
        One can clearly see that the amount of dense material that has accumulated around Sgr A* is different in both maps.
        Although in the model WR\_f07 (see bottom left-hand side panel of Figure~\ref{fig:rho_maps}) some dense material is spiraling towards the black hole overall there are is no clear structure around it. 
        In the case of WR\_f1 (see bottom right-hand side panel of Figure~\ref{fig:rho_maps}), the dense material has settled at the centre in a a disc-like structure. 
        This difference is due to the efficiency of the radiative cooling. 
        Since model WR\_f1 has enhanced cooling, more dense clumps and filaments form, and some of them manage to fall onto Sgr~A*. 
        Overall, the hydrodynamic evolution of the two models is analogous: the winds fill the domain during the first $\sim$500~yr, then the systems reach a quasi-steady state, and in the last 500~yr they diverge as the model WR\_f1 creates much more dense, cool material that settles around Sgr~A*. 
        A quantitative analysis of the evolution of the properties of system is presented as follows.

        To quantify the accretion rate at different spatial scales we calculated the mass flux as a function of both radial distance and time $\dot{M}(r,t)=4\pi r^2 \rho(r,t)v_r(r,t)$ averaged over a spherical shell. 
        Here, positive and negatives signs in the radial velocity refer to outflow and inflow mass fluxes, respectively. 
        First, we analyse the net mass flux at the innermost radius we can resolve properly, which is equivalent to five times the inner boundary radius, i.e. $5r_\text{in}=5\times10^{-4}~\text{pc}\approx 1.25\times10^{-2}~\arcsec$. 
        Figure~\ref{fig:mdot} shows $|\dot{M}(r=5r_\text{in},t)|$ as a function of time for models WR\_f07 and WR\_f1 that are displayed as solid blue and orange lines, respectively. 
        In both cases, at $t>-3300~\text{yr}$ the net mass flux reaches levels of $\sim$10$^{-6}~\text{M}_{\odot}~\text{yr}^{-1}$ with short variability episodes that increase the rate by factors of two to four. 
        During this quasi-steady phase, both simulations display similar behaviour qualitatively, likely determined by the identical stellar wind configuration. 
        This stage lasts until $t\approx-500~\text{yr}$ where the mass-flow rates start to deviate from each other. 
        The fiducial model continues to display a variability amplitude of the same order of magnitude.
        However, the WR\_f1 model shows a transition to a strongly gas inflow dominated phase, with inflow rates that are enhanced by a factor four to eight. 
        This stage of the evolution is what we refer to as the disc formation phase, analogously to our previous models reported in \cite{calderon2020}. 
        The spatial distribution of the inflowing material is not simple. 
        The fiducial case shows the same behaviours described in previous work \citep[e.g.][]{ressler2018} where there are two components: one that most of the time follows the same orientation of the stars in the clockwise disc, and other that comes from the south pole of such a disc. 
        In the enhanced case, the behaviour is analogous but the inflow component along the disc is more relevant. 
        In both cases, the inflowing material exhibits a significant range in angular momentum orientation as discussed further in this section. 
        Thus, none of them shows a simple thin disc accretion scenario but rather a thick disc with an extra component along the south pole.

        Next, we proceed to analyse the time-averaged behaviour of the simulations over the last $500~\text{yr}$, as this can give us an idea of the general state of the system minimising the effects of the stochastic variability. 
        First, we analyse the (volume-weighted) density and (mass-weighted) temperature radial profiles of the simulations, which are shown on the top and bottom panels in Figure~\ref{fig:prof_rhoT}, respectively. 
        The dashed blue and solid orange lines represent the simulations WR\_f07 and WR\_f1, respectively. 
        The orange lines in both panels clearly highlight the presence of the cold disc. 
        The density profile decays with $r^2$ in both cases at large scales ($\gtrsim1\arcsec$). 
        This is the result of most of the material flowing outwards at these scales following a roughly isotropic spherical wind as seen in previous models \citep{ressler2018,calderon2020}. 
        However, the profiles differ at smaller scales ($<1\arcsec$) where the fiducial case transitions to $\rho\propto r^{-1}$, while the model WR\_f1 shows a density enhancement due to the presence of the disc. 
        This increase in density is more than one order of magnitude. 
        Regarding the temperature profiles, both models match at larger scales ($\gtrsim1\arcsec$) with a constant temperature of the order of $10^7~\text{K}$, set by the stellar wind collisions. 
        Again, the profiles differ at smaller scales where the fiducial case follows $T\propto r^{-1}$, and the enhanced cooling case displays a temperature profile about two orders of magnitudes lower on average. 
        The minimum in temperature corresponds to the region where the disc contributes with most of the mass for a given spherical shell.

        To quantify and characterise the mass inflow and outflow regimes in the simulation domain we calculated them as radial profiles. 
        Figure~\ref{fig:mdot_profile} displays the absolute value of the mass flow rates as a function of distance from Sgr~A*. 
        The solid and dashed lines represent the mass inflow and outflow rates, respectively; while the colours follow the same convention as before.
        Thus, if the solid line is above the dashed line the net mass flow is inwards and vice-versa. 
        Analogous to the model by \cite{ressler2018}, the fiducial model encompasses different regimes due to the dominant direction of the mass flow at given spatial scales. 
        At $r>3\arcsec$, the outflow component dominates and the inflow is negligible. 
        We find a net mass outflow rate of $\sim5\times10^{-4}~\text{M}_{\odot}~\text{yr}^{-1}$. 
        Notice that the enhanced cooling model exhibits exactly the same behaviour at these scales. 
        At smaller scales ($1\arcsec-3\arcsec$), there is a transition region where $|\dot{M}_\text{in}|\sim|\dot{M}_\text{out}|$ that corresponds to the location of most mass-losing stars and wind interactions. 
        For model WR\_f1, the change is more abrupt due to the presence of the cold disc. 
        In both models, at $r<1\arcsec$ the inflow component is larger than the outflow, so the material inflows across these scales and down to the innermost boundary. 
        The net effect makes the mass inflow rate about five times larger which generates the cold disc.

        The origin of the infalling material can be traced in two ways: analysing the angular momentum of the material close to the inner boundary, and using the scalar tracer fields that we introduced to label the chemical abundances of the different WR subtypes. 
        Figure~\ref{fig:hammer} shows Hammer projections of the angular momentum direction of the gas enclosed in a sphere of radius $0.25\arcsec$ ($\sim0.01~\text{pc}$) for the fiducial and enhanced cooling runs in the top and bottom panels, respectively. 
        Each point represents a different simulation time over the last 1000~yr, and they are connected with dashed lines.
        For reference, we added the angular momentum direction of the orbits of the mass-losing stars as black star symbols, and the location of the clockwise disc based on the orbits we used as a grey shaded circle. 
        This analysis shows that the angular momentum direction of the infalling material varies stochastically but overall tends to align with the orientation of the orbits in the clockwise disc. 
        However, the degree of variability depends on whether or not the model results in the formation of a disc. 
        The fiducial model shows more variability while the enhanced cooling run displays that the angular momentum aligns more consistently with the orientation of the clockwise disc. 
        The fact that the infalling material at this spatial scales comes from these stars agrees with previous numerical models \citep{cuadra2008,ressler2018,calderon2020}. 
        Furthermore, this is not surprising since these stars orbit closer to Sgr~A*, and have winds that are relatively slow ($\sim600$~kms), which results into shocks more prone to radiative cooling and with smaller angular momentum. 
        It is also relevant to mention that the average angular momentum direction of the accreted material has a significant dispersion, despite being similar to the direction of the angular momentum of the stars in the clockwise disc. 
        Although not shown here, the angular momentum direction distribution shows a polar thickness of $\sim$40$^{\circ}$ for the bulk of the inflow material, and in some cases $\sim$100$^{\circ}$. 
        This means that the inflow takes place preferentially along the orientation of the clockwise disc (or the cold disc) but also along higher inclinations close to the disc orientation. 
        Such a behaviour can be seen in both the fiducial and enhanced models, and are consistent with previous work results \citep[e.g.][]{ressler2018}. 
        
        Figure~\ref{fig:prof_tracer} shows radial profiles of the scalar tracers. 
        The dashed, solid, and solid-dashed represent the scalars that correspond to the WR subtypes WC89, WN57, and WN89/Ofpe, respectively. 
        The blue and orange lines stand for the results for the models WR\_f07 and WR\_f1. 
        In this analysis, the scalar tracer values represent the fraction of the mass density that comes from a given WR subtype wind. 
        Thus, in general most of the material supplied into the domain comes from the WN89/Ofpe stars. 
        At smaller scales ($r<1\arcsec$), the dominance is even clearer as the material from these stars makes about 80-90\% of the mass budget. 
        This is also the result of these stars having slow winds and colder shocked material. 
        The rest of the stars contribute mostly to the outflow of material, as their contribution is more relevant at larger scales. 
        Notice that both models, fiducial and enhanced, display roughly the same behaviour across the entire domain.

        Overall, the fiducial model displays properties of the medium and gas dynamics consistent with the previous models that do not form a disc  \citep[run for 1100~yr]{cuadra2008,ressler2018,ressler2020, calderon2020}. 
        The model with enhanced cooling is consistent with the simulations that show the formation of a cold disc as an outcome of this system \citep{calderon2020}. 
        Before discussing if any of the models can be favoured given the current observational constraints we proceed to characterise the cold disc arising in the WR\_f1, as its properties will aid us to do so.

    \subsection{Properties of the cold disc}

        At the present time ($t=0$), the model WR\_f1 shows the presence of a cold disc. 
        Figure~\ref{fig:disc_maps} shows zoomed maps of the central region of $2.5\arcsec\times$~$2.5\arcsec$. 
        The left- and right-hand side panels show projected maps of density and line-of-sight velocity (both weighted by density), respectively. 
        Here it is possible to observe that the projected diameter of the disc is roughly $1\arcsec$. 
        We estimated the mass of the disc by isolating the cold material ($T<10^5~\text{K}$) and integrating the cell mass content within a sphere of radius 1.0\arcsec. 
        We found that the total mass of the disc is 0.005~M$_{\odot}$ which is consistent with the value in our previous work \citep{calderon2020}.
        However, the mass accumulated in the disc does not converge and keeps increasing at least for the next hundreds of years in the simulation. 
        
        On the right-hand side panel of Figure~\ref{fig:disc_maps}, we can see that the line-of-sight velocity peaks at $\sim$2000~km~s$^{-1}$ along both directions (towards and outwards the observed). 
        This coincides with the maximum extension of the observed H30$\alpha$ line \citep{murchikova2019}. 
        Additionally, the disc is observed to be $\sim45^{\circ}$ tilted in projection. 
        This is not in agreement with the observations since, at least in projection the difference is about $\sim90^{\circ}$ \cite[see Figure~1 of][]{murchikova2019}. 

        \begin{figure*}
            \centering
            \includegraphics[width=0.495\textwidth]{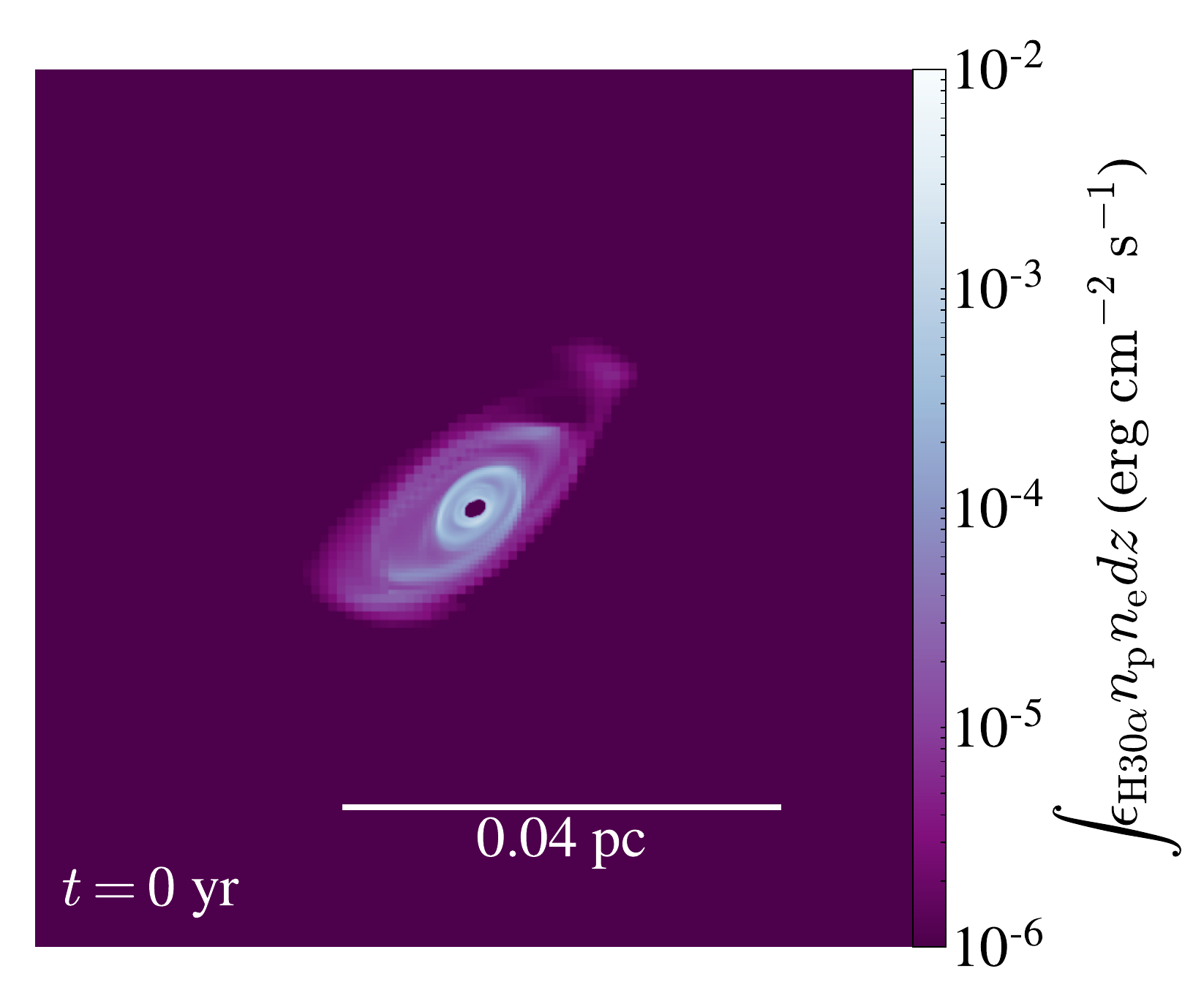}
            \includegraphics[width=0.495\textwidth]{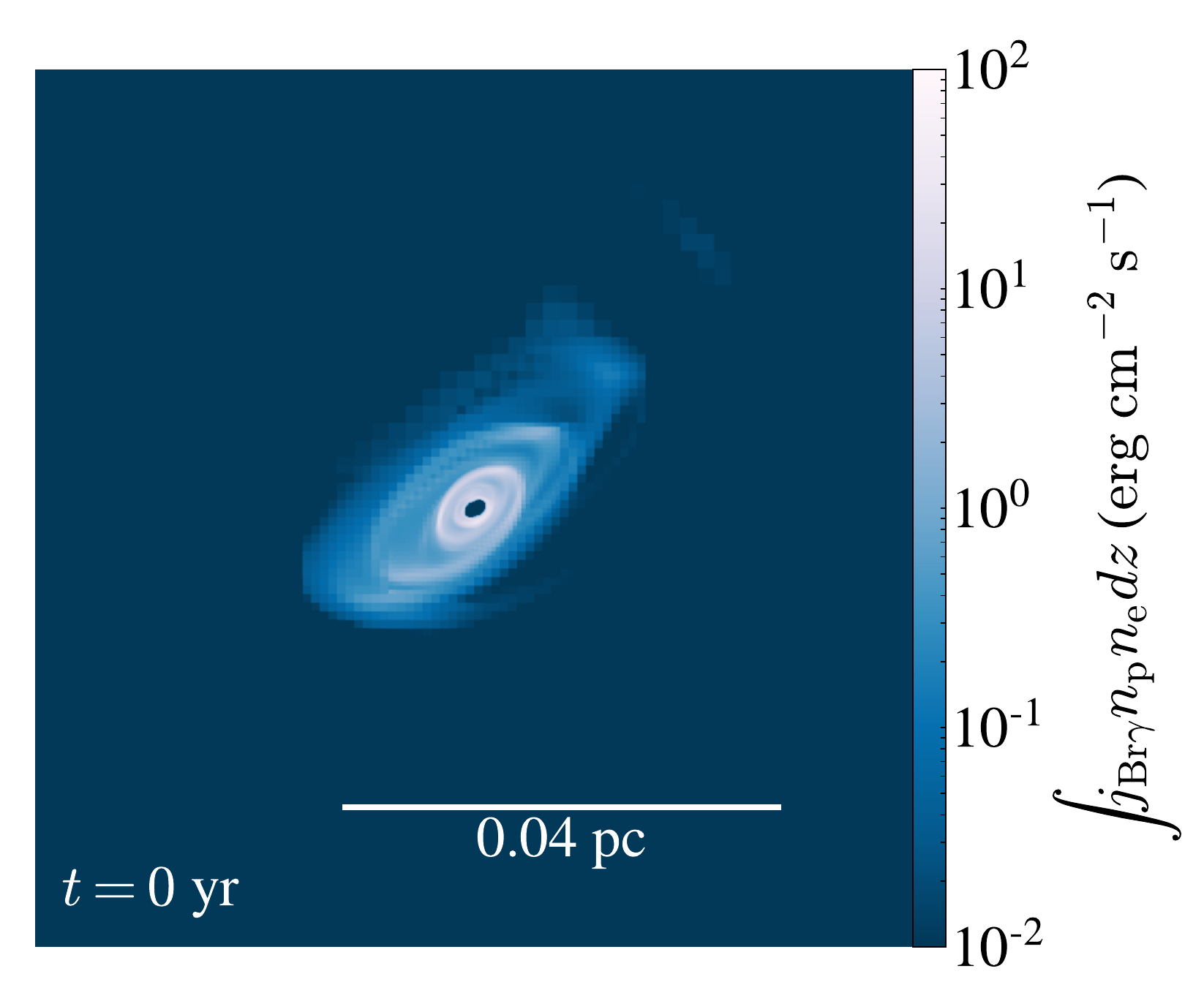}
            \caption{
            Line-of-sight integrated maps of the emission of the H30$\alpha$ and Br$\gamma$ recombination lines shown on the left- and right-hand side panels.
            The maps correspond to the inner 2\arcsec$\times~$2\arcsec~of the model WR\_f1. 
            }
            \label{fig:line_maps}
        \end{figure*}

    \section{Post-processing of observables}
    \label{sec:mock}

        In order to assess the validity of our models for the Galactic centre we proceed to synthesise observational quantities that we could compare easily with observations. 
        We focus on the recombination lines H30$\alpha$ and Br$\gamma$. 
        Afterwards, we calculate the X-ray spectrum from our models. 

    \subsection{Recombination lines: H30$\alpha$ and Br$\gamma$}
    \label{sec:lines}        
        The cold gas at the location of Sgr~A* has been observed through the H30$\alpha$ recombination line \citep{murchikova2019,yusef-zadeh2020}, and only with an upper limit in the Br$\gamma$ recombination line \citep{ciurlo2021}. 
        In order to explore if the models are consistent with these observations we have calculated the expected flux from these emission lines. 
        To compute the emission of the H30$\alpha$ line we used the coefficients given below fitting a piecewise linear function to them as a function of density and considering a decay $\propto T^{-1}$ \citep[see][supplementary information]{murchikova2019},
        \begin{eqnarray}
            \epsilon_{\text{H}30\alpha}(10^4~\text{K},10^4~\text{cm}^{-3})
            &=&
            1.05\times10^{-31}~\text{erg}~\text{s}^{-1}~\text{cm}^3,
            \label{eq:h30a_1}\\
            \epsilon_{\text{H}30\alpha}(10^4~\text{K},10^5~\text{cm}^{-3})
            &=&
            1.08\times10^{-31}~\text{erg}~\text{s}^{-1}~\text{cm}^3,
            \\
            \epsilon_{\text{H}30\alpha}(10^4~\text{K},10^6~\text{cm}^{-3})
            &=&
            1.25\times10^{-31}~\text{erg}~\text{s}^{-1}~\text{cm}^3,
            \\
            \epsilon_{\text{H}30\alpha}(10^4~\text{K},10^7~\text{cm}^{-3})
            &=&
            1.36\times10^{-31}~\text{erg}~\text{s}^{-1}~\text{cm}^3.
            \label{eq:h30a_4}
        \end{eqnarray}
        In the case of the Br$\gamma$ line, its emissivity can be calculated following \cite{schartmann2015} as
        \begin{equation}
            j_{\text{Br}{\gamma}}=3.44\times10^{-27}\left(\frac{T}{10^4~\text{K}}\right)^{-1.09}~\text{erg}~\text{s}^{-1}~\text{cm}^3.
            \label{eq:hbrg}
        \end{equation}
        Then, in each cell of the domain we computed the respective emissivity values, multiplied them by their $n_\text{p}n_\text{e}$ (assuming full ionisation), and then integrated the data cube along the $z$ coordinate which is perpendicular to the line of sight. 
        It is important to remark that in this calculation it is necessary to take into account the hydrogen nucleus $n_\text{p}$ and electron $n_\text{e}$ densities appropriately. 
        Since our models consider different amounts of hydrogen in the plasma we include only the material that contains some, i.e. the material coming from the WN89/Ofpe stars.
        Additionally, we only used the hydrogen fraction of that material which is $X_\text{H}=11.5\%$ and $X_\text{H}=40\%$ for the models WR\_f07 and WR\_f1.
        
        Figure~\ref{fig:line_maps} shows the maps over a region of 2\arcsec$\times$~2\arcsec centred in Sgr~A* that resulted from this procedure. 
        Although the maps were also calculated for the model WR\_f07 we chose not to show them here since the emission is negligible (four orders of magnitude smaller) due to the lack of cold gas that contributes to the recombination line flux.  
        From these maps we estimated the flux as seen from Earth, that is, at a distance of 8.33~kpc, so that they could be compared directly with the observed values. 
        These were calculated by integrating within a circular area of a projected radius  $p=0.23\arcsec$ and $p=1\arcsec$ centred at location of Sgr~A*. 
        These radii were motivated by the observations that used an aperture of 0.23\arcsec for extracting the flux observed . 
        However, our models show that the disc actually extends beyond this size, so to get an idea of the whole flux of the model we chose a size of $p=1\arcsec$ that encloses entirely the disc.
        The flux values obtained are shown in Table~\ref{tab:fluxes} where we also added the upper limit for Br$\gamma$ from \cite{ciurlo2021}, and the flux of H30$\alpha$ measured from integrating the spectrum across the velocity channels given by \cite{murchikova2019}. 
        In the case of the Br$\gamma$ line, we can see that only the model WR\_f07 has a flux consistent with the upper limit reported for an aperture of $p=0.23\arcsec$. 
        The model WR\_f1 has a flux that is $\sim100$ times higher than the upper limit. 
        Also, we see that the disc in our models extends beyond this projected radius, yet its emission is dominated by the inner $p=0.23\arcsec$ (see Table~\ref{tab:fluxes}). 
        The situation is more complex for the H30$\alpha$ line since the observed flux is in both cases higher than the synthetic emission. 
        Although the model that forms a disc is closer it still remains 30\% times fainter than the reported value. 
        The model WR\_f07 has negligible emission as it is ten order of magnitudes fainter compared to the observation. 
        Thus, it is still not clear how to reconcile the models with both line emission simultaneously as none of them is capable to reproduce the observations. 
        We discuss further the interpretation of these results in Section~\ref{sec:disc}.
        \begin{table*}
            \centering
            \begin{threeparttable}
            \caption{Radiative flux of the recombination lines from central region.}
            \begin{tabular}{lcccc}
                \hline
                Source &
                $f_{\text{Br}\gamma}(p<1\arcsec)$ & $f_{\text{H}30\alpha}(p<1\arcsec)$ & $f_{\text{Br}\gamma}(p<0.23\arcsec)$ & $f_{\text{H}30\alpha}(p<0.23\arcsec)$\\
                (1)&(2)&(3)&(4)&(5)
                \\
                \hline
                \hline
                &&&&
                \\
                Observed & - & - & $\leq7.26\times10^{-15}$ & $(3.0\pm0.6)\times10^{-17}$
                \\
                WR\_f07 & $6.08\times10^{-17}$& $6.70\times10^{-22}$ & $1.18\times10^{-18}$ & $1.34\times10^{-27}$
                \\
                WR\_f1& $5.74\times10^{-13}$&
                $2.19\times10^{-17}$& $5.32\times10^{-13}$& $2.05\times10^{-17}$
                \\
                \hline
            \end{tabular}
            \label{tab:fluxes}
            \begin{tablenotes}
                \item
                \textit{Notes.} 
                All fluxes correspond to quantities seen from Earth, i.e. at a distance of 8.33~kpc, and are given in units erg~cm$^{-2}$~s$^{-1}$. 
                The observed H30$\alpha$ flux was calculated integrating the reported flux density across the velocity channels, and taking into consideration the error propagation of the 0.3~mJy reported \citep[see Figure~1 in][]{murchikova2019}. 
            \end{tablenotes}
            \end{threeparttable}
        \end{table*}

        \begin{figure}
            \centering
            \includegraphics[width=0.495\textwidth]{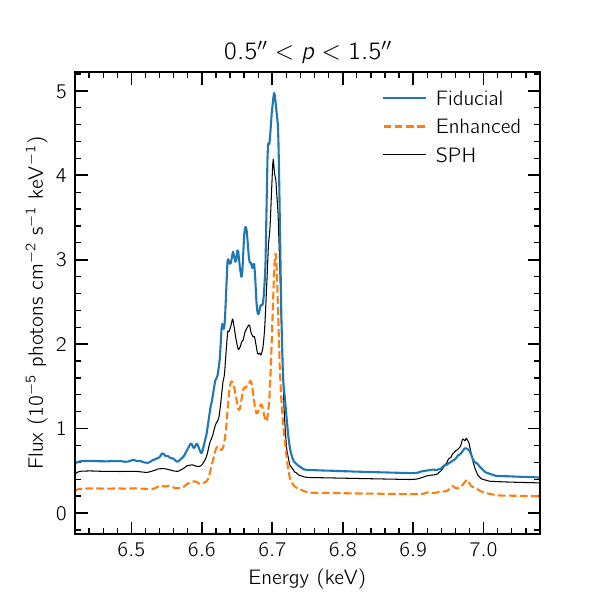}
            \caption{
            Synthetic X-ray spectra computed from the numerical models in an annulus of 0.5\arcsec$<p<$1.5\arcsec. 
            The models WR\_f07 and WR\_f1 are shown in solid orange and dashed blue lines, respectively. 
            Additionally, the thin solid black line corresponds to the results from the SPH model \citep{balakrishnan2024b}.
            }
            \label{fig:spectrum}
        \end{figure}

        \begin{figure*}
            \centering
            \includegraphics[width=0.495\textwidth]{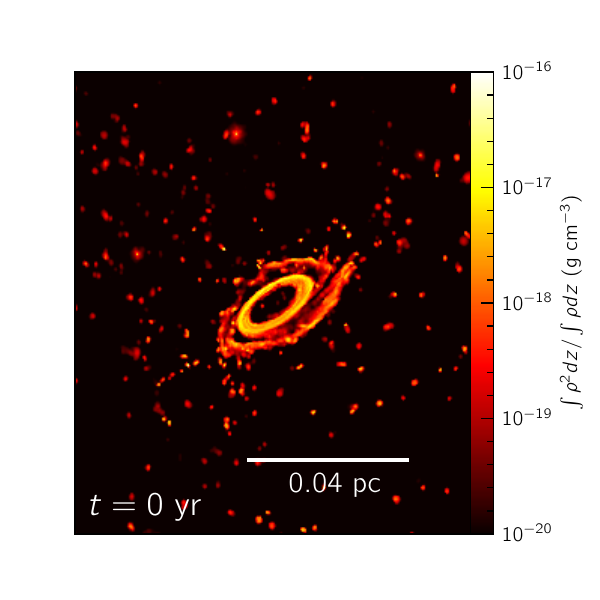}\includegraphics[width=0.49\textwidth]{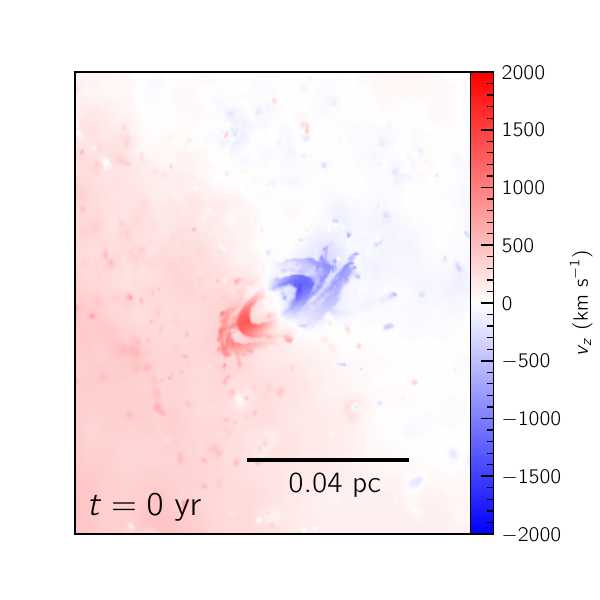}
            \caption{
            Density and line-of-sight velocity maps of the central 2.5\arcsec$\times$~2.5\arcsec of the SPH model at $t=0$. 
            The left- and right-hand side panels show projected density and line-of-sight velocity, respectively. 
            These panels are analogous to the ones shown in Figure~\ref{fig:disc_maps}.}
            \label{fig:sph_maps}
        \end{figure*}

    \subsection{Spectral X-ray emission}
    \label{sec:xray}

        The X-ray emission from the central parsec is also an observable that has allowed validation of previous numerical models \citep[see][]{russell2017,ressler2018,wang2020}. 
        Here we also post-processed our models aiming to obtain the Iron K-alpha spectrum of this region at $\sim6.7$~keV as this is the strongest X-ray feature from Sgr~A* \citep[e.g.][]{russell2017,corrales2020}. 
        Our approach is based on the procedure outlined in \cite{balakrishnan2024b} but adapted to our finite-volume grid-based hydrodynamic model, which is briefly described as follows. 
        First, we assigned energy dependent X-ray emissivities  $j^k_E = n^k_e n^k_i\Lambda(E,T^k) $ that were taken from the \textsc{vvapec} model \citep{smith2001}, which was obtained with \textsc{xspec} \citep{arnaud1996}. 
        The densities and temperatures were taken from the last output of our simulation, i.e. at $t=0$. 
        Additionally, the tracer fields were used to identify the composition fraction of the parcels of gas according to the WR subtypes. 
        Since the optical depth through the simulation domain is optically thin in the X-ray, the radiative transfer equation to solve is simply given by the integration along the line of sight of the emissivities of the respective cells. 
        Given that we can trace the fraction of the material that comes from a given WR subtype this actually corresponds to a linear combination of the tracer field $s_i$ and the abundance dependent emissivity for each cell, i.e.  
        \begin{eqnarray}
            I_E(x,y)
            &=&
            \sum_{i=1}^3 s_i\int j^k_E(x,y,X_i)dz
            \\
            &=&
            \sum_{i=1}^3 s_i\int n^k_\text{ion}n^k_e\Lambda(E^k,T_i,X_i)\phi^k(v_\text{los})dz,
        \end{eqnarray}
        where $\phi^k$ represents a Gaussian function that it is used to obtain the emission across different velocity channels, and to take into account the thermal broadening represented with a velocity of $v_\text{th}$. 
        This function is given as follows
        \begin{equation}
            \phi^k(v_\text{los})=\frac{1}{\sqrt{2\pi}v_\text{th}}\exp\left\{-\frac{(v^k_\text{los}-v_\text{los})^2}{2v_\text{th}^2}\right\}.
        \end{equation}
        We performed calculations through velocity channels spanning line-of-sight velocities $v_\text{los}$ from  -3000~km~s$^{-1}$ to 3000~km~s$^{-1}$. 
        These were later converted into energy space via Doppler broadening. 
        Thus, for each energy bin from the emissivity tables we obtained 150 line profiles using a fine resolution of 6400 per dex. 
        In order to sum them appropriately we interpolate them into a unique energy range so that we can obtain a single spectrum that includes both the continuum and line contributions. 
        The result of this procedure is shown in Figure~\ref{fig:spectrum}. 
        This contains the X-ray spectrum emitted from a sky-projected annulus of $0.5\arcsec<p<1.5\arcsec$ for models WR\_f07 (solid blue line) and WR\_f1 (dashed orange line). 
        As a reference, we included the estimation from \cite{balakrishnan2024b} computed from an analogous model that used the SPH approach, which we discuss in the following section.
        All spectra show qualitative agreement highlighting the different features of the Fe complex at 6.7~keV and 6.9~keV. 
        This is also reflected in the broadening of the lines due to the velocity across the line of sight, which is expected since the models are based on the same stellar orbits and wind properties.
        
        It is relevant to highlight the level of agreement between the finite-volume and SPH models. 
        The qualitative agreement is reasonable and the level of the emission is of the same order of magnitude. 
        The exact quantitative difference among the models could be mainly attributed to the exact cooling table (composition) used in each simulation, which is the result of the different density and temperature distributions that can be observed in Figure~\ref{fig:prof_rhoT}. 
        This is why the WR\_f07 shows the highest flux level as it has the least efficient cooling and, as a result it has hotter material. 
        Analogously, the run WR\_f1 with an enhanced cooling specified in the composition of its plasma results in a reduction of $\sim$40\% of its flux.
        In case of the SPH model, the quantitative differences could be affected due to the intrinsic ability of the generic SPH to capture shocks which impacts directly the density, velocity structure, and the temperature of the medium. 
        Notice that the flux level of the spectrum of the SPH model by \cite{balakrishnan2024b} is within the cases explored in this work. 
        Their cooling model is not the same as the ones explored here but this analysis show that it is equivalent to a cooling efficiency between both of our models. 

        Although not shown here we also compared the X-ray spectra from simulation WR\_f1 before and after the disc is formed, and found negligible differences.
        Finally, we also analysed the slope of the continuum emission within this energy range and found no significant differences among the models. 
        Thus, in this work the presence of the disc does not imprint a feature in the X-ray spectrum that is resolvable by any current or forthcoming X-ray telescope, in agreement with \cite{balakrishnan2024b}.

\section{Comparison with Lagrangian models}
\label{sec:sph}

    Hydrodynamic simulations of the feeding of Sgr~A* by the WR stellar winds have been conducted with different numerical tools. 
    In this work, we have presented the results of using a finite-volume approach that solves the hydrodynamic equations in the Eulerian form. 
    However, there has been extensive work on this problem using codes that solve the equations in the Lagrangian form. 
    Specifically, the use of SPH has been the main choice for such a task \citep{cuadra2005,cuadra2006,cuadra2008,cuadra2015,russell2017,wang2020}. 
    Despite the limitations of the generic SPH technique (e.g. capturing shocks) the models managed to reproduce the observed accretion rate at the Bondi radius \citep{cuadra2008} as well as the X-ray emission \citep{russell2017}. 
    In this context, we have conducted a comparison of the Eulerian models calculated with Ramses, and the Lagrangian models computed with the SPH code Gadget \citep{springel2005}. 
    The SPH simulation setup is as similar as it can be to our setup. 
    This corresponds to an updated version from the models in \cite{russell2017}. 
    For more details on the setup of the SPH models, we refer the reader to \cite{balakrishnan2024a,balakrishnan2024b}. 
    We focus the comparison on the analysis on the final state of the system, specifically in the cold discs formed after 3000~yr.
    
    Figure~\ref{fig:sph_maps} shows density and line-of-sight velocity maps projected along the line of sight of the SPH model at $t=0$ on the left- and right-hand side panels, respectively.
    Notice that these panels are analogous to the ones shown in Figure~\ref{fig:disc_maps} for the Eulerian models presented in this work. 
    First, let us compare the disc density maps between the two approaches. 
    Simply through visual inspection it is possible to see that the structure of the discs is not exactly the same. 
    The Eulerian disc has a more continuous structure towards smaller scales while the Lagrangian disc displays a ring-like structure.
    This feature can be attributed to the spatial resolution and the inner boundary radius. 
    The Eulerian models indeed manage to resolve smaller scales towards the centre, and the radius of the inner boundary is $5~\text{mas}$ while the Lagrangian models resolve $\sim$1~mas and have an inner boundary of $0.1\arcsec$. 
    As a result, the Lagrangian model obtained a disc 50 times lighter ($\sim10^{-4}~\text{M}_{\odot}$) than in the Eulerian simulation. 
    This difference could be attributed to the radiative cooling employed in each simulation: WR subtype abundance and Solar with 3$Z_{\odot}$ in the Eulerian and the Lagrangian model, respectively. 
    Regarding the projected orientation of the disc both models seem to be in agreement. 
    This can be analysed more carefully on the line-of-sight velocity maps (see right-hand side panels of Figures~\ref{fig:disc_maps} and~\ref{fig:sph_maps}). 
    Both maps display that the discs indeed match their projected orientation as well as their most blue- and redshifted velocities are roughly -2000~km~s$^{-1}$ and 2000~km~s$^{-1}$. 

    Overall, the agreement on the outcome a cold disc around Sgr~A* whose properties align between two different approaches indicates that this is a result independent of the numerical technique. 
    Thus, this shows that the determinant factors to obtain this results lies in both the long-enough simulation time ($\gtrsim3000~\text{yr}$ ) as well as efficient radiative cooling. 

\section{Disc stability analysis}
\label{sec:stability}
   In this section, we present an analysis of the stability of the observed cold disc in the Galactic Centre.
   Based on its reported properties, the disc should be depleted by accretion after its viscous timescale $t_{\nu}\approx43~\text{kyr}$, assuming an $\alpha=0.1$ thin disc \citep{shakura1973}, a process much slower than its generation as modelled in our simulations, although faster than its formation out of CND material \citep{solanki2023}. 
   It is important to bear in mind that our hydrodynamic simulations do not take into account physical viscosity therefore only numerical viscosity is present. 
   Nonetheless, there are several external mechanisms that could be capable of destroying the disc faster than through accretion. 
   Among them are the gravitational and hydrodynamic interactions of the disc with the stars, the stellar-wind--disc collisions due to the presence of the S-star cluster in its location, and/or the evaporation due to heat flowing between the hot medium and the cold disc.
   Although some of these physical scenarios have been studied analytically and numerically, there has not been any direct application to the cold disc discovered in our Galactic Centre.

    \subsection{Stellar cluster perturbing a thin disc}
    \label{sec:grav}
    
       The stability of a cold disc perturbed by stellar passages has been investigated by \cite{ostriker1983}. 
       Their work presented an analytical approach to calculate how the disc loses angular momentum due to the passages, considering the integrated effect of a stellar cluster in stationary state.
       Besides this work, most efforts have been put on following how the stellar dynamics are affected by the disc presence \citep[e.g.][and references therein]{fabj2020} or on the effect of \emph{embedded} accreting stars and/or black holes on the disc \citep[e.g.][]{takawa2021} rather than on the consequences to the disc properties. 
       
       In this section, we summarise the approach of \cite{ostriker1983} and apply it to the Galactic centre case.
       Let us start considering a cold, thin accretion disc around a compact object. 
       Within the thin disc we assume that its height $H(r)$ is small compared to the radial coordinate $r$, i.e. $H(r)/r\ll1$. 
       Also, we consider that the sound speed of the material in the disc is small compared to the speed of the stars at the same radius. 
       In this scenario, a star passing through the disc will produce a torque that removes angular momentum from the disc. 
       We consider that the star interacts with the disc gravitationally and hydrodynamically 
       The gravitational interaction of a star crossing the disc can be described using the Bondi-Hoyle approach \citep{bondi_hoyle1944}, while the hydrodynamic interaction is just a physical collision of a solid sphere with a gaseous disc.
        
        Let us consider a system of reference where the $x$ axis is in the direction of rotation and $z$ is in the direction of the rotation pole of the disc. 
        Then, according to \cite{ostriker1983} the total change of linear momentum along the $x$ direction $\Delta p_x$ for a single stellar passage is   
         \begin{equation}
            \Delta p_x=-\pi R_*^2\Sigma_\text{d}\left[q+\ln\Lambda_{\rm D}\eta^4\left(\frac{v_0}{v_{\rm rel}}\right)^4\right]\left(\frac{v\cos\phi-v_{\rm d}}{\sin\theta\cos\phi}\right)\left(\frac{v_{\rm rel}}{v}\right),
         \end{equation}
         where $R_*$ is the radius of the star, $\Sigma_\text{d}=\int_\text{disc}\rho_\text{d} H(r)$ is the surface density of the disc, $\rho_\text{d}$ is the volume density of the disc, $q$ is the hydrodynamic drag parameter that is set to $q=2$ assuming a large Mach number collision, $v_0$ is the local stellar velocity,  $\Lambda_\text{D}\approx H(r)v_0^2/(R_*v_*^2)$ is the Coulomb logarithm, $v_*$ is the escape velocity from the surface of the star, $\eta=v_*/v_0$ is the hardness parameter, $v_\text{d}$ is the disc velocity, $v_\text{rel}$ is the relative velocity vector, and $(\theta,\phi)$ are the spherical coordinate angles along the polar and azimuthal directions. 
         
         In order to consider the effect of the complete stellar cluster interacting with the disc we need to integrate over the cluster phase-space volume. 
         First, we integrate over the local stellar velocity distribution. 
         Let us define $dN$ as the number of stars in the velocity range $v\to v+dv$, i.e   
         \begin{equation}
            dN=4\pi n_*v_0^{-3}f(v)v^2dv, 
         \end{equation}
         where $f(v)$ is the velocity distribution function, $n_*$ is the stellar number density.
         Then, the number of crossings per unit time and area in the $(\theta,\phi)$ direction is
         \begin{equation}
            dN_\text{cr} = n_*v_0^{-3}f(v)v^2dv\sin\theta d\theta d\phi\sin\theta\cos\phi.
         \end{equation}
         
         Now, integrating over the phase space, the total transfer of momentum to the disc per unit area at $r$ is
         \begin{eqnarray}
            \dot{p}_{x,\text{tot}}(r)&=&-2\pi^2R_*v_0^{-3}\Sigma_\text{d}(r)n_*(r)\int^{\infty}_0dvv^2f(v)\nonumber\\
            &&\times
            \int^{+1}_{-1}d\mu(v\mu-v_\text{d})v_\text{rel}\left[q+\eta^4\ln\Lambda_\text{D}(v/v_\text{rel})^4\right],\nonumber\\
            &&
         \end{eqnarray}
         where $\mu=\cos\phi$.
        
         As the local momentum per unit area in the disc is $p_{x,\text{d}}(r)=\Sigma(r)/\tau_\text{d}(r)$ the inverse of the drag timescale to remove the angular momentum of the disc is $\tau_\text{d}=-p_{x,\text{d}}/\dot{p}_{x,\text{tot}}$ can be written as
            \begin{equation}
                \tau_\text{d}=\left[n_*R_*^2v_0\left(qI_0+\eta^4\ln\Lambda_\text{D}I_1\right)\right]^{-1},
                \label{eq:ostriker}
            \end{equation}
            where the quantities $(I_0,I_1)$ depend solely on the stellar distribution function. 
            For instance, in the case of a Maxwellian distribution the values of these parameters can be obtained numerically and are $I_0=12.553$ and $I_1=0.474$.

            In order to apply this model to the Galactic Centre we need to calculate the stellar density within the size of the disc.  
            Following \cite{schodel2007} the stellar mass density at $r<6~\arcs$ is described by 
            \begin{equation}
                \rho_*(r)=2.8\pm1.3\times10^7~\msun~\text{pc}^{-3}\left(\frac{r}{6~\arcs}\right)^{-1.2}.
            \end{equation}
            Assuming that the stellar mass density is composed mainly by stars whose radii are $R_*=1~\rsun$ and move on their orbits typically at $v_0=1000~\kms$ we can use Equation~\ref{eq:ostriker} to estimate the inverse of the drag timescale,
            \begin{eqnarray}
               \tau_\text{d} & = & 185\left(\frac{n_*}{3.8\times10^8~\text{pc}^{-3}}\right)^{-1}\left(\frac{R_*}{1\rsun}\right)^{-2}\left(\frac{v_0}{10^3~\kms}\right)^{-1}\nonumber\\ %25.106
               &\times&\left[1+0.08\left(\frac{v_*}{440~\kms}\right)^4\left(\frac{v_0}{10^3~\kms}\right)^{-4}\left(\frac{\ln\Lambda_\text{D}}{10}\right)\right]^{-1}\text{Myr}.\nonumber\\
               &&
            \end{eqnarray}
            Based on this calculation the angular momentum of the cold disc should be removed after $\sim185~\text{Myr}$ of interactions of the stars. 
            This is much longer than the formation time-scale found in our simulations.
            Furthermore, it is even much longer than the age of the Wolf-Rayet stars, which is about $6~\text{Myr}$ \citep{martins2007}. 
            Then, the gravitational and hydrodynamic interactions of the stars onto the disc should not affect the stability of the disc.   

         \subsection{Stellar winds perturbing a disc}
         \label{sec:wind}

            In the case of stellar irradiation affecting the state of the accretion flow there have been many studies mainly motivated by the recent pericentre passages of the star S2 around Sgr~A*.
            For instance, \cite{cuadra2003} could rule out the existence of an optically thick disc based on the lack of its thermally reprocessed emission as it gets illuminated by S2. 
            \cite{nayakshin2004} estimated that a star as luminous as S2 could heat up and ionise an inner disc, which could enhance the accretion rate.
            \cite{giannios2013, christie2016} calculated the bremsstrahlung emission as a result of the stellar wind shocking a Radiatively Inefficient Accretion Flow (RIAF) around Sgr~A*, finding an X-ray luminosity comparable to quiescent emission of Sgr~A*. 
            In fact, no noticeable increase was detected during 2018 \citep[see Table~1 of][]{andres2022}. 
            \cite{hosseini2020} constrained the observed $L'$-band variability of S2 to be about 2--3\%, based on a model of bow shock of its stellar wind, implying an ambient density $<10^5\,$cm$^{-3}$, which rules out the presence of a standard accretion disc at S2's pericentre, in agreement with the previous studies. 
            It is important to remark however that such a limit marginally allows the existence of both the disc reported by \cite{murchikova2019}, and also the one in our simulation. 

            The stars in the S cluster also have relatively powerful stellar winds. 
            As the stars inhabit the black hole in the vicinity of the cold disc, it is expected that the winds interact with it. 
            By either opening a bubble due to the high kinetic energy carried in the winds or depositing thermal energy they affect the disc properties. 
            We proceed to perform analytical estimates to quantify the impact of these processes. 
    
            \subsubsection{Bubbles in the disc}
    
               Let us consider a star blowing an isotropic stellar wind with a mass-loss rate $\dot{M}_\text{w}$ at a terminal speed $v_{\rm w}$. 
               Then, its density $\rho_\text{w}$ at a distance $r$ from the star is given by
               \begin{equation}
                  \rho_{\rm w}(r)=\frac{\dot{M}_\text{w}}{4\pi r^2v_{\rm w}}
               \end{equation}
               If such a star is immersed in a medium whose number density is $n_\text{m}$ at rest at a temperature $T_\text{m}$, the radius of the bubble $r_\text{b}$ opened by the wind can be calculated equating the ram pressure of the wind and the thermal energy of the medium, i.e.
               \begin{eqnarray}
                   r_\text{b} & = &1.48\times10^{15}\left(\frac{\dot{M}_\text{w}}{10^{-8}~\msunyr}\right)^{1/2}\left(\frac{v_{\rm w}}{1000~\kms}\right)^{1/2}\nonumber\\
                   &&\times\left(\frac{n_{\rm m}}{10^6~\cc}\right)^{-1/2}\left(\frac{T_{\rm m}}{10^4~\kelvin}\right)^{-1/2}~\text{cm},
               \end{eqnarray}
               where we have used as fiducial values the disc parameters measured by \cite{murchikova2019} and typical stellar wind properties for B stars \citep{oskinova2011,krticka2014}.
               Then, the radius of the bubble is about 0.012~\arcs, which is approximately one tenth of the disc radius. 
       
               In order to open such a bubble a star would need to be within the disc for at least the wind crossing time, which is the minimum time it takes for a bubble of that size, i.e. $r_\text{b}/v_{\rm w}\approx0.5~\text{yr}$. 
               Let us estimate the duration of a stellar passage through the disc.
               Along the perpendicular direction a star crossing the disc would take at most $0.1R_\text{d}/v_0$, being $H(r)/r=0.1$. 
               One S star moves typically at $v_0\approx1000~\kms$, then $0.1R_\text{d}/v_0\approx1~\yr$. 
               As both timescales are comparable, we conclude that the bubble can be opened.
            
               For such a perturbation to last we need to check if either shear or sound waves could eliminate it, i.e. close the bubble. 
               In the case of shear, this is given by the orbital speed of the disc and the size of the bubble:
               \begin{equation}
                  t_\text{close}=\frac{r_\text{b}}{\Delta v_\text{kep}}=\sqrt{\frac{r}{GM}}\left[r+r_\text{b}+\sqrt{r(r+r_\text{b})}\right],
               \end{equation}
               where $\Delta v_\text{kep}$ corresponds to the difference in Keplerian velocity at the centre and edge of the bubble. 
               As this expression increases with $r$, the largest value will be at the outer edge of the disc, then $t_\text{close}<5~\text{yr}$. 
               This value should be interpreted as the longest duration of a perturbation of the wind onto the disc. 
               As there are 12 S-stars on orbits whose pericentre distances are shorter than the radius of the disc and their orbital periods are of the order of $\sim10~\text{yr}$ \citep{gillessen2017} we expect a bubble to be opened every year on average. 
               However, as shear would close such bubbles in less than five years there would be at most four bubbles opened on average in a stationary state. 
               As the size of the bubbles is very small compared to the size of the disc (two orders of magnitude in area), four of them would not have an impact on its structure. 
               The sound crossing timescale $r_\text{b}/c_\text{s}\approx 30~\text{yr}$ is longer than the shear timescale so we do not expect it to have an impact on erasing perturbations.

            \subsubsection{Winds injecting thermal energy}
               
               The S stars are of spectral type B with masses of 10~\msun\  \citep[e.g.][]{habibi2017}, and thus their mass-loss rate must be $\sim10^{-8}~\msunyr$ with terminal velocity of 1000~\kms. 
               The supersonic nature of the winds should be enough to compress the shocked material at high temperature.
               Potentially, this thermal energy could be deposited onto the disc during each stellar passage. 
               If we consider that most of the kinetic energy of the wind is transformed into thermal energy about $\sim6.4\times10^{33}~\erg$ would be deposited in the disc during the $\sim 1\,$yr long passage (see above). 
               In order to check if the wind energy can significantly impact the thermodynamic state of the disc let us estimate its total internal energy $U_\text{int}$. 
               \begin{equation}
                  U_\text{int}=\frac{n_\text{d}k_\text{B}T_\text{d}}{\gamma -1}V_\text{d},
               \end{equation}
               where $k_\text{B}$ is the Boltzmann constant and $V_\text{d}$ is the disc volume. 
               Replacing the observed disc properties we obtain that $U_\text{int}=1.63\times10^{42}~\erg$. 
               From this calculation it is clear that even if all the energy from the wind in a stellar passage is deposited the contribution is still negligible to modify the state of the disc. 
               Even if we consider the integrated effect of all the S stars passages on a timescale comparable to the viscous timescale of the disc ($\sim43~\text{kyr}$) the contribution is still $<0.1\%$ of the total thermal energy of the disc.
               Thus, the energy supplied by the stellar winds into the disc is not enough to affect its state.
    
   \subsection{Thermal conduction}
   \label{sec:conduction}

      The vicinity of Sgr~A* at the Bondi radius is made out of a diffuse ($10~\cc$), hot medium \citep[$10^7~\text{K}$;][]{baganoff2003} due to the shocked stellar winds from the Wolf-Rayet stars in the region. 
      At the disc vicinity ($\sim10^3\ R_\text{Sch}$), the conditions are more extreme as the plasma has a density of about $\sim5\times10^3~\cc$ \citep{gillessen2019} and temperature of $\sim10^8~\kelvin$. 
      As the disc is significantly colder it is expected that heat flows from the medium to the disc. 
      The large temperature gradient should enable thermal conduction to take place and likely evaporate the disc.

      The problem of evaporating a cold structure sitting in a hot environment due to thermal conduction was first studied by \cite{cowie1977}. 
      In that work, the authors derived an analytical expression for the mass-loss rate and evaporation timescale for a spherically symmetric gas cloud. 
      Unfortunately, these expressions are not valid for more complex geometries.
      \cite{meyer1994} studied specifically the evaporation of a cold thin accretion disc in a hot corona. 
      Although the model considered only one radial zone it was able to calculate the full perpendicular structure of the accretion disc and the corona by means of numerical simulations. 

      \cite{liu2004} used the model by \cite{meyer1994} to analyse which solutions were consistent with the observational data available at the time in case there were a cold disc in the center of the Galaxy. 
      They found that any transient disc would evaporate quickly. 
      With their obtained evaporation rate of $\sim 10^{-4}\,\msunyr$, the life-time of a disc with properties such as found by \cite{murchikova2019} or formed in our models would be $\sim 1\,$yr. 
      Similar analysis have argued that small clumps ($\sim$M$_{\oplus}$) in the Galactic Centre should also evaporate quickly ($\sim$1~yr) due to thermal conduction \citep{burkert2012,calderon2018}. 
      According to this, cold gas should not be able to last long if formed, yet we observe many cold gas structures in the central parsec: the many dusty cold clumps detected in the vicinity of IRS~13E \citep{fritz2010,peissker2023}, and a larger gaseous cloud X7 \citep{ciurlo2023}. 
      Thus, this regime of classical thermal conduction might not be applicable to the Galactic centre environment. 
      Another possibility has been studied by \cite{nayakshin2004b} who focused on the same problem and found solutions in a different regime, where the electron mean free path is long enough that an accretion disc does not evaporate but rather increases its mass by condensation.         
      A more thorough analysis of the role of thermal conduction is beyond the scope of this work and is deferred to future study.

      Our findings shows that gravitational and hydrodynamic effects require timescales that are too long to have an impact on the disc stability. 
      On the other hand, we find that thermal conduction in the classical case should evaporate the observed disc in about one year. 
      Given the time-scale for disc formation found in our model, the \cite{nayakshin2004b} condensation regime remains to our knowledge as the only viable physical scenario that would allow its continued existence and potential identification with \cite{murchikova2019}'s reported disc.

\section{Discussion}
\label{sec:discussion}

    In this section, we interpret our findings exploring the uncertainties in the abundances of WR stars in general and of the ones in the central parsec. 
    Additionally, we contrast the properties of the simulated and observed cold discs as well as discuss further implications.

    \subsection{Is the simulated disc the observed disc?}
    \label{sec:disc}

        In order to compare the simulated and the observed discs, first we analyse their physical properties.         
        The simulated disc has a total mass of $\sim5\times10^{-3}~\msun$, while the structure observed in H30$\alpha$ was reported to have a mass of $10^{-5}-10^{-4}~\msun$. 
        However, comparing these quantities might not be appropriate since the mass of the observed structure is based on three assumptions: a thin disc with uniform density and temperature, and a masing factor of $\sim100$. 
        All of them minimise the value of the mass inferred from the observations. 
        For instance, dropping the maser assumption the inferred mass increases in a factor ten \citep[see Equations 30-31 of Supplementary Material in][]{murchikova2019}. 
        Bearing this in mind, we could argue that the mass of the simulated and observed disc might be of the same order provided there was no maser. 
        Although removing that assumption would create tension reconciling both the Br$\gamma$ and H30$\alpha$ observed fluxes. 
        
        Regarding its extension, the simulated disc has a projected radius of $\sim0.5\arcsec$, while the observed disc has a radius of $0.23\arcsec$. 
        However, since most of the recombination line emission of the disc is originated from its central region ($p<0.23$\arcsec, see Table~\ref{tab:fluxes}) the sizes of the simulated and observed discs are in agreement. 

        A more direct comparison can be done through the analysis of the mock observational quantities computed from the simulations. 
        In Section~\ref{sec:lines}, we computed the H30$\alpha$ and Br$\gamma$ recombination line emission and estimated their fluxes. 
        Here, we have found that both synthetic fluxes are in tension with the observations. 
        First, the simulated Br$\gamma$ flux is 100 times higher than the upper limit. 
        It is relevant to remark that this limit is already extinction corrected. 
        Second, the simulated H30$\alpha$ flux is 30\% lower than the observed one. 
        These results differ from the estimates by \cite{ciurlo2021} from our previous models due to, again, the uniform disc assumptions. 
        The simulated disc is not thin or possess uniform density or temperature. 
        Although the mean density and temperature of the simulated disc is of the order of $\sim10^{5}~\text{cm}^{-3}$ and $\sim10^4~\text{K}$ its geometry is more complex with a larger variance in its properties. 
        As a result, both the synthetic Br$\gamma$ and H30$\alpha$ fluxes are much higher than the ones inferred by \cite{ciurlo2021}. 
        
        Based on the line-of-sight velocity maps shown in \cite{murchikova2019}, the orientation of the observed disc displays the red- and blueshifted sides to be on the East (left-hand) and West (right-hand) sides, respectively. 
        Although our model WR\_f1 also shows this configuration the exact inclination of the discs does not match perfectly. 
        Specifically, the discs are inclined in $\sim90^{\circ}$ among each other, being the simulated disc tilted in the clockwise direction on the sky. 
        In principle, this argues against the WR stars as the main responsible for the formation of the observed disc. 
        But we cannot rule out completely their role in the process, since a combination of the WR stars with other gaseous structures, such as the minispiral or the CND, acting simultaneously might result in a net effect that manages to reproduce the observed angular momentum. 
        \cite{solanki2023} showed that material from the circumnuclear disc could fall close to Sgr~A* though on much longer timescales. 
        Certainly, this scenario would be much more difficult and challenging to simulate with the appropriate chemical abundances as well as with high-enough spatial resolution. 

        It is also relevant to notice that the observed accretion flow orientation at $\lesssim10~\text{R}_\text{Sch}$, where $\text{R}_\text{Sch}$ corresponds to the Schwarzschild radius of Sgr~A*, is consistent with the orientation of the clockwise stellar disc \citep{baubock2020,akiyama2022}. 
        Thus, our models reproduce that connection between the dynamics from large ($>10^5~\text{R}_\text{Sch}$) to small ($\lesssim10~\text{R}_\text{Sch}$) scales. 
        However, the observed cold disc does not seem to follow such a connection.

        Finally, an intriguing aspect of our findings is that the WR atmospheres with more H in them would favour the formation of the disc. 
        At the same time, the disc indeed was detected in the H$30\alpha$ recombination line, which obviously indicates the presence of H. 
        Thus, it is sensible to ask how much H can be in the winds of WR stars but, specifically in the ones of the subtype WN89/Ofpe.  
        This point is discussed in the following section.
        
        \subsection{Uncertainties in abundances}
        \label{sec:H}

           The results of the hydrodynamic models demonstrate that the chemical abundances of the material in the WR winds are the key factor to determining whether or not a cold disc can be formed due to their action. 
           Unfortunately, there are two large uncertainties in this context. 
           First, the metallicity of the young population of the nuclear star cluster is not well constrained \cite[e.g.][]{genzel2010}. 
           Although there is agreement that it is above Solar its exact value has not been established so far, with most studies quoting the range $Z=2-3Z_{\odot}$. 
           However, even if the metallicity is determined more precisely, the problem would not be entirely solved, since the abundances of the plasma in the region are dominated by the material supplied by the WR star winds whose compositions are much more uncertain. 
           Theoretically, the evolution of these stars is a topic of active research, and improved prescriptions for mass loss are changing the modelled properties of observed WR stars \citep[see e.g.][and references therein]{gormaz2023}.
           Simply based on observations, up to now, it has been reported that some of these stars might possess from $X_\text{H}=0$ to $X_\text{H}=50\%$ for Galactic WR stars of the WN subtype \citep{hamann2006,hamann2019}, and even as high as $X_\text{H}=60\%$ for extragalactic ones \citep{todt2015}.
           In the vicinity of Sgr~A*, \cite{martins2007} characterised the wind properties of the sample of WR stars through the analysis of their infrared spectra. 
           They reported abundance H/He ratios in the range 2--5 for these stars which does not allow us to rule out the scenario regarding this aspect. 
           A variation on the hydrogen abundance in the chemical mixture in a factor five might change the net radiative cooling rate by a factor $1.1-1.4$, especially at lower temperatures. 
           In our models, WR\_f07 considered a fiducial value of hydrogen mass fraction of 11.5\% \citep{russell2017} but the model WR\_f1 assumed 40\%. 
           Such a change was enough to modify the cooling function by about 30\%, which ended up affecting significantly the final result and the state of the plasma. 
           Thus, within our current understanding of the abundances in the WR stellar atmospheres both scenario are plausible. 
           Only better quality spectra and more sophisticated models of them could allow us to constrain the wind properties and their composition more precisely.

\section{Conclusions}
\label{sec:conclusions}

   We present a numerical study of the system of mass-losing stars feeding Sgr~A*, focusing on the impact of the chemical composition of the plasma. 
   Through studying simulations with different abundances, we find that the system can either form or not form a cold disc around the black hole, provided that the model is run for sufficient timescales ($\gtrsim3000~\text{yr}$) to create a dense enough medium in the inner parsec of the Galaxy.
   As in our previous work, we confirm that, if formed, the cold disc can significantly
impact the hydrodynamic and thermodynamic state of the plasma within the central parsec. 
   As a result, the inferred mass-inflow rate at $5\times10^{-4}~\text{pc}$ can reach up to $8\times10^{-6}~\text{M}_{\odot}~\text{yr}^{-1}$, which is between four and eight times higher than the value without the presence of such a disc. 
   Our models still point to the mass-losing stars in the clockwise disc being the main source of the accreted material due to their dense and slow winds, as previous models have shown \citep[e.g.][]{cuadra2008,ressler2018,calderon2020}. 
   In this work, we have followed the origin of the inflowing material more carefully through the use of passive scalar fields for different WR subtypes. 
   This allowed us to confirm the origin of the accreted material but more importantly led us to conclude that this is mostly provided by the winds of the stars of the WR subtype WN89/Ofpe. 
   This provides evidence that the infalling material must have a non-negligible fraction of hydrogen, as these WR stars have not entirely lost their atmospheric hydrogen via winds. 

   Our models also show that the formation of a cold disc depends strictly on the radiative cooling employed, which is determined by the chemical abundances of the plasma. 
   Previous models only considered solar abundances with $3Z_{\odot}$ but do not agree on whether or not a cold disc can be formed. 
   From our results in Appendix~\ref{app:solar}, we speculate that at $Z\approx 3Z_{\odot}$ there is a transition between the regimes of persistent quasi-steady accretion and disc formation, which would imply that small differences in the cooling function or numerical implementation can result in models reproducing either regime. This may explain the differing results obtained by \cite{calderon2020} and \cite{ressler2020} with similar numerical and physical setups.
   
   In the case where a cold disc is formed around Sgr~A*, the structure may resemble  the extension and the line-of-sight velocity structure of the observed disc. 
   However, we are not able to reproduce direct observational quantities. 
   Specifically, the exact sky-projected inclination of the simulated and observed discs differs by $~\sim90^{\circ}$. 
   The upper limit on the Br$\gamma$ emission line is 100 lower than the synthetic flux.  
   The observed H$30\alpha$ emission line flux is 30\% higher than in our model. 

   Furthermore, we contrasted the Eulerian (finite-volume grid-based) models of  this work with analogous (SPH) Lagrangian models in order to address the potential influence of the chosen approach on the outcome of the simulations. 
   \cite{balakrishnan2024b} showed that a cold disc is also formed around Sgr~A* if the simulation is run for long timescales. 
   Both cold structures agree on their sizes, their velocity structure, and their sky-projected inclination. 
   Although there are differences in the total mass and specific inner structure of the discs, these could be attributed to the resolution employed and the exact assumption to consider the innermost boundary. 
   Despite these differences, the agreement is reasonable among the numerical approaches, especially when analysing the synthetic X-ray emission (see Figure~\ref{fig:spectrum}). 
   The spectral features agree qualitatively across the models presented in this work and the Lagrangian model. 
   The levels of the X-ray emission are all within the same order of magnitude but the exact base level depends on the radiative cooling chosen, which is determined by the chemical abundances in the WR atmospheres. 
    
    We have also explored the stability and long-term evolution of the putative disc under the effect of perturbing agents: the nuclear star cluster, the wind of all the mass-losing stars (including the S stars) impacting the disc through opening bubbles and/or depositing thermal energy via shocks, and the potential effect of thermal conduction. 
    Our analysis shows that gravitational and hydrodynamic effects are likely to impact the disc on longer timescales than the viscous timescale. 
    However, within the classical thermal conduction framework, a cold disc in such a region should not exist. 
    Based on this and the disc-formation timescale from our simulations, we speculate that the condensation model of \cite{nayakshin2004b} remains a plausible mechanism consistent with the existence of a disc. 

    In conclusion, through the present work, we have been able to reconcile the discrepancy among the numerical simulations of the system of mass-losing stars feeding Sgr~A*. 
    The formation of a cold disc on a $\sim$3000~yr timescale is indeed possible for certain chemical abundances that are consistent with the current observational constraints. 
    Nonetheless, it is not possible to favour or disfavour this scenario due to the related uncertainties. 
    Thus, high-resolution spectra of the WR stars and more sophisticated modelling of them could allow us to discern the validity of such scenarios. 
    We must warn that even in the case where the cold disc is a result of the action of the WR stars, it is not possible to reproduce all of its observed properties, specifically its inclination and recombination line fluxes. 
    More sophisticated models that include infalling material from other sources might be necessary to produce the exact inclination of the structure, which could also impact the emission of the structure. 
    
\begin{acknowledgements}
    We thank Prof. Sergei Nayakshin and Prof. Stanley Owocki for very helpful discussions about this project. 
    We also thank Dr. Sean M. Ressler for sharing the radiative cooling function shown in Figure~\ref{fig:lambdas}. 
    Additionally, we are grateful to Dr. Álex Gormaz-Matamala for helping with references and discussions on the WR wind abundances. 
    DC and JC acknowledge the support of the Kavli Foundation through its summer program at the Max Planck Institute for Astrophysics where fruitful discussions took place.
    In addition, DC thanks the warm hospitality of the Universidad Adolfo Iba\~nez in Chile and University of Delaware in the USA where part of this project was developed. 
    We acknowledge the support from ANID in Chile through FONDECYT Regular grant 1211429 and Millennium Science Initiative Program NCN$2023\_002$.
    The research of DC and SR was supported by the Deutsche Forschungsgemeinschaft (DFG, German Research Foundation) under Germany’s Excellence Strategy - EXC 2121 - ``Quantum Universe” - 390833306. 
    Since 01.10.2024, DC has been funded by the Alexander von Humboldt Foundation. 
    C.M.P.R. acknowledges support from NASA Chandra Theory grant number TM3-24001X. Resources supporting this work's SPH simulations were provided by the NASA High-End Computing (HEC) Program through the NASA Advanced Supercomputing (NAS) Division at Ames Research Center.
    SR is also supported by the European Research Council (ERC) Advanced Grant INSPIRATION under the European Union’s Horizon 2020 research and innovation programme (Grant agreement No. 101053985), the Swedish Research Council (VR) under grant number 2020-05044, and the research environment grant ``Gravitational Radiation and Electromagnetic Astrophysical Transients” (GREAT) funded by the Swedish Research Council (VR) under Dnr 2016-06012, by the Knut and Alice Wallenberg Foundation under grant Dnr. KAW 2019.0112.  
    The numerical simulations of this work were run on the high-performance computing system \textsc{cobra} of the Max Planck Computing and Data Facility. 
    Most of the analysis of the numerical data was carried out using the \textsc{python} package \textsc{yt} \citep{turk2011}.
    Furthermore, this work made use of \textsc{python} libraries \textsc{numpy} \citep{harris2020} and \textsc{matplotlib} \citep{hunter2007}, as well as of the NASA’s Astrophysics Data System.
\end{acknowledgements}
   \bibliographystyle{aa} % style aa.bst
   \bibliography{gc-disc} % your references Yourfile.bib
\begin{appendix}
    
    \section{Numerical setup choices}
    \label{app:solver}

       \cite{calderon2020} simulated the system of WR stars feeding Sgr~A* while orbiting around it following an analogous numerical setup compared to the models presented in this work. 
       There, we had found that the resulting cold disc tended to align with the grid after its formation. 
       In order to remedy this numerical artifact we investigated the use of a MinMod slope limiter, instead of MonCen. 
       However, we warned that it was not straightforward to maintain the spherical symmetry of the winds using the MinMod slope limiter. 
       In this work, we investigated more carefully these numerical choices in order to be confident that our results are robust.

       First, we investigated the choice of the MonCen slope limiter. 
       To do this, we made use of the same setup presented in this work but paying attention to the values of the passive scalar fields. 
       In principle, the values of these fields should be between zero and one. 
       However, this experiment showed that the MonCen slope limiter produced non-physical oscillations that in some cases resulted in negative values of the passive scalar fields. 
       This behaviour was not observed when using the MinMod slope limiter. 
       Thus, we selected this choice throughout this work. 
       To ensure the spherical symmetry of the stellar winds we had to impose a less restrictive refinement criterion based on density gradients as well as a smoother transition between refinement levels through the parameter \textsc{nexpand}. 

\begin{figure*}
            \centering
            \includegraphics[width=0.6\textwidth]{plots/colorbar.pdf}
            \includegraphics[width=0.475\textwidth]{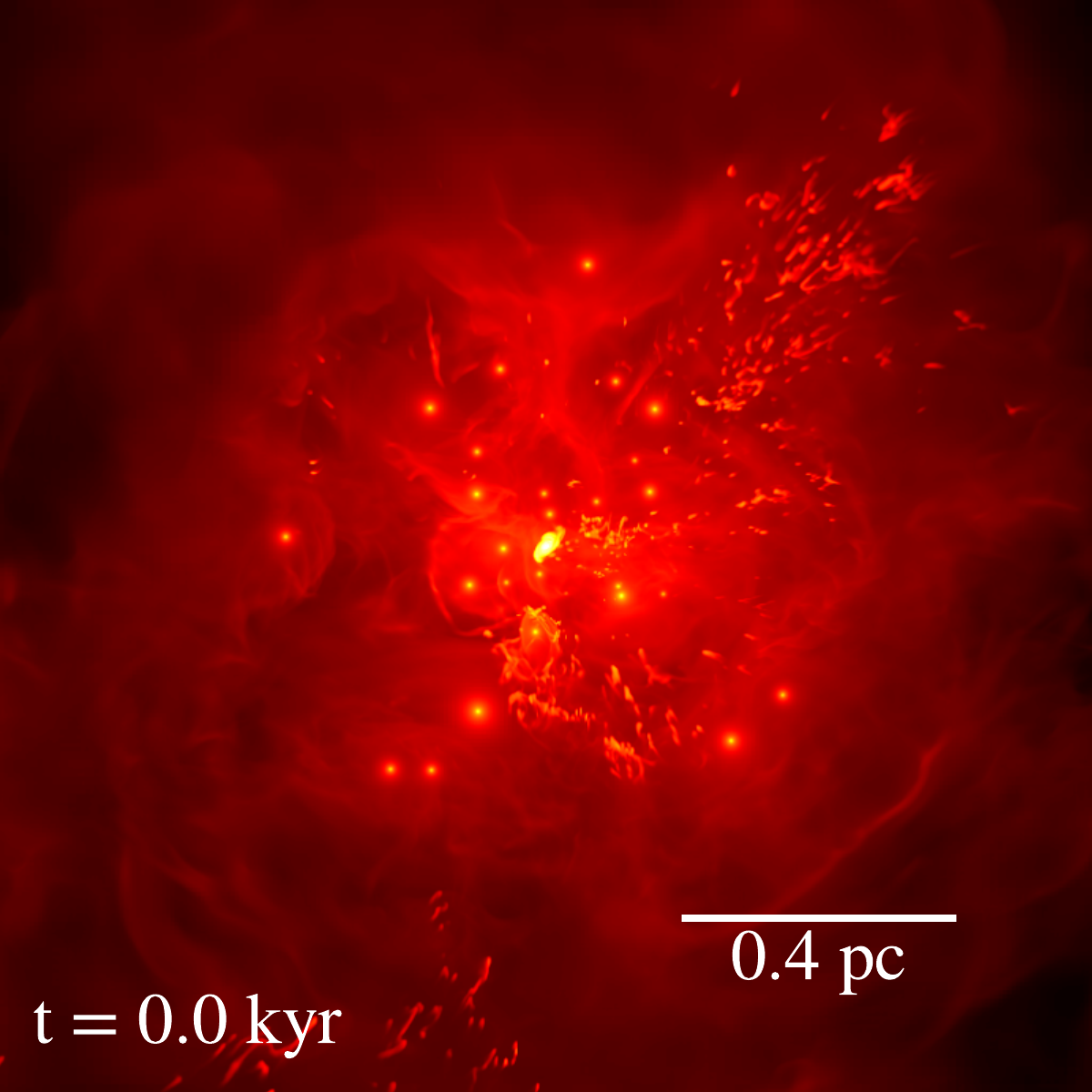}
            \hspace{0.01cm} \includegraphics[width=0.475\textwidth]{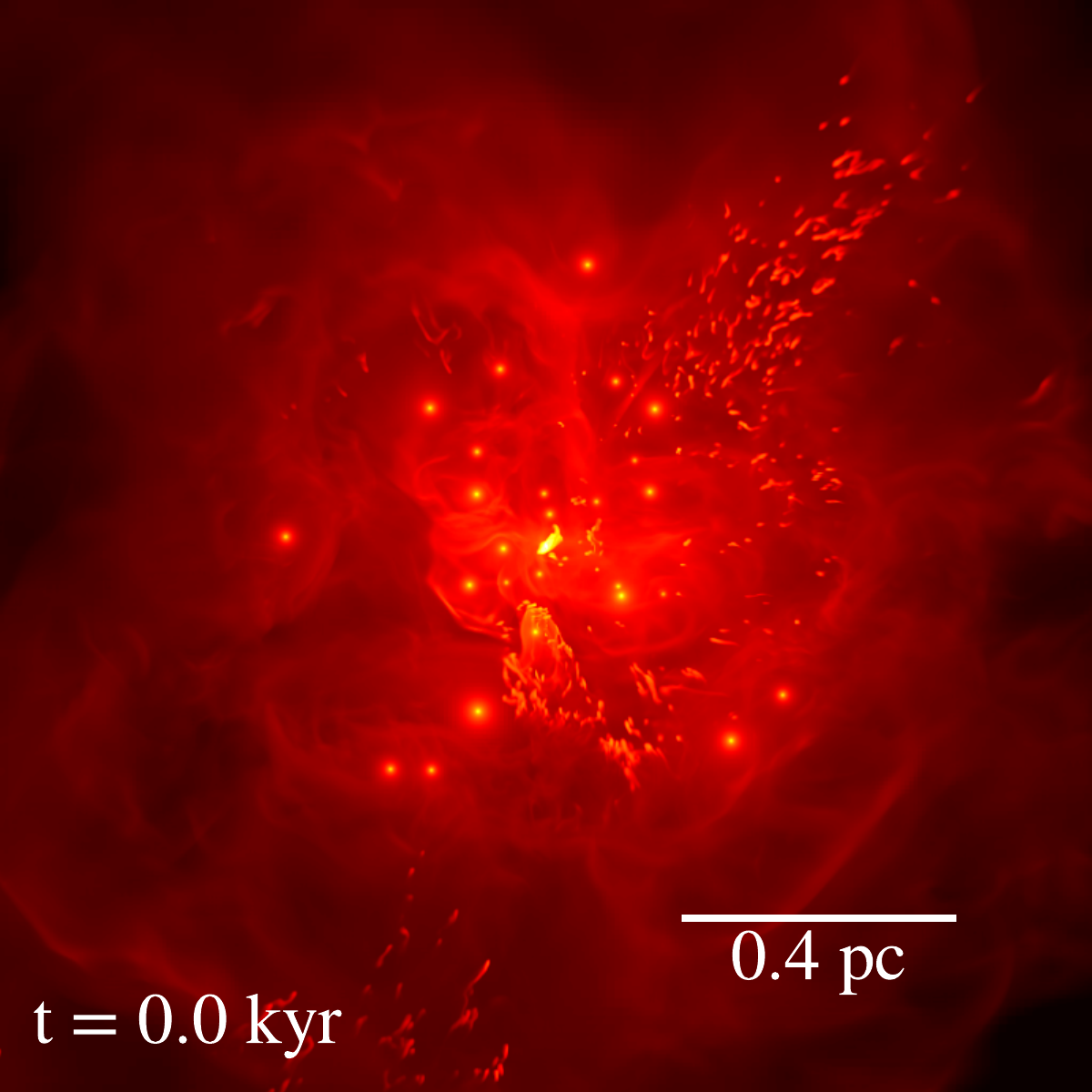}
            \caption{Comparison of the final state (present time) of the simulations with different Riemann solvers. 
            Left- and right-hand side panels correspond to runs with Harten-Lax-van Leer-Contact (HLLC) and exact Riemann solvers, respectively.
            Both models were run using the MinMod slope limiter.
            }
            \label{fig:comparison}
        \end{figure*}

       Second, we tested the choice of the Riemann solver employed but using the MinMod slope limiter.
       Currently, the HLLC Riemann solver \citep{toro1994} is widely used in grid-based hydrodynamic simulations. 
       In this work, we have found that this choice results in a cold disc that tends to align with the Cartesian grid as seen in \cite{calderon2020}. 
       This could indicate problems in conserving the angular momentum which is a known issue with simulations in Cartesian grids when the relevant regions are not well resolved. 
       With the usage of the \textit{exact} Riemann solver \citep{toro2009}, though more computationally expensive, it was possible to avoid this artificial alignment.
       Thus, the only sensible combination was the use of the \textit{exact} Riemann solver with a MinMod slope limiter but we had to ensure to count with enough resolution and smoothness in the transition of the refinement levels. 
       As a reference of these experiments, Figure~\ref{fig:comparison} shows density projection (weighted by density) maps of the final state of the system for two runs with Solar composition with $3Z_{\odot}$ but with different Riemann solvers. 
       The left- and right-hand side panels display the models using the HLLC and the \textit{exact} Riemann solvers, respectively. 
       Although overall the density structure looks similar there are subtle differences. 
       The most relevant is related to the disc orientation and structure. 
       On the left-hand side, the disc tries to align with one of the Cartesian axes, while on the right-hand side the disc remains consistently with the shown orientation.

    \newpage
    \section{Solar composition varying metallicity}
    \label{app:solar}

\begin{figure*}
            \centering
            \includegraphics[width=0.6\linewidth]{plots/colorbar.pdf}
            \\
            \includegraphics[width=0.325\linewidth]{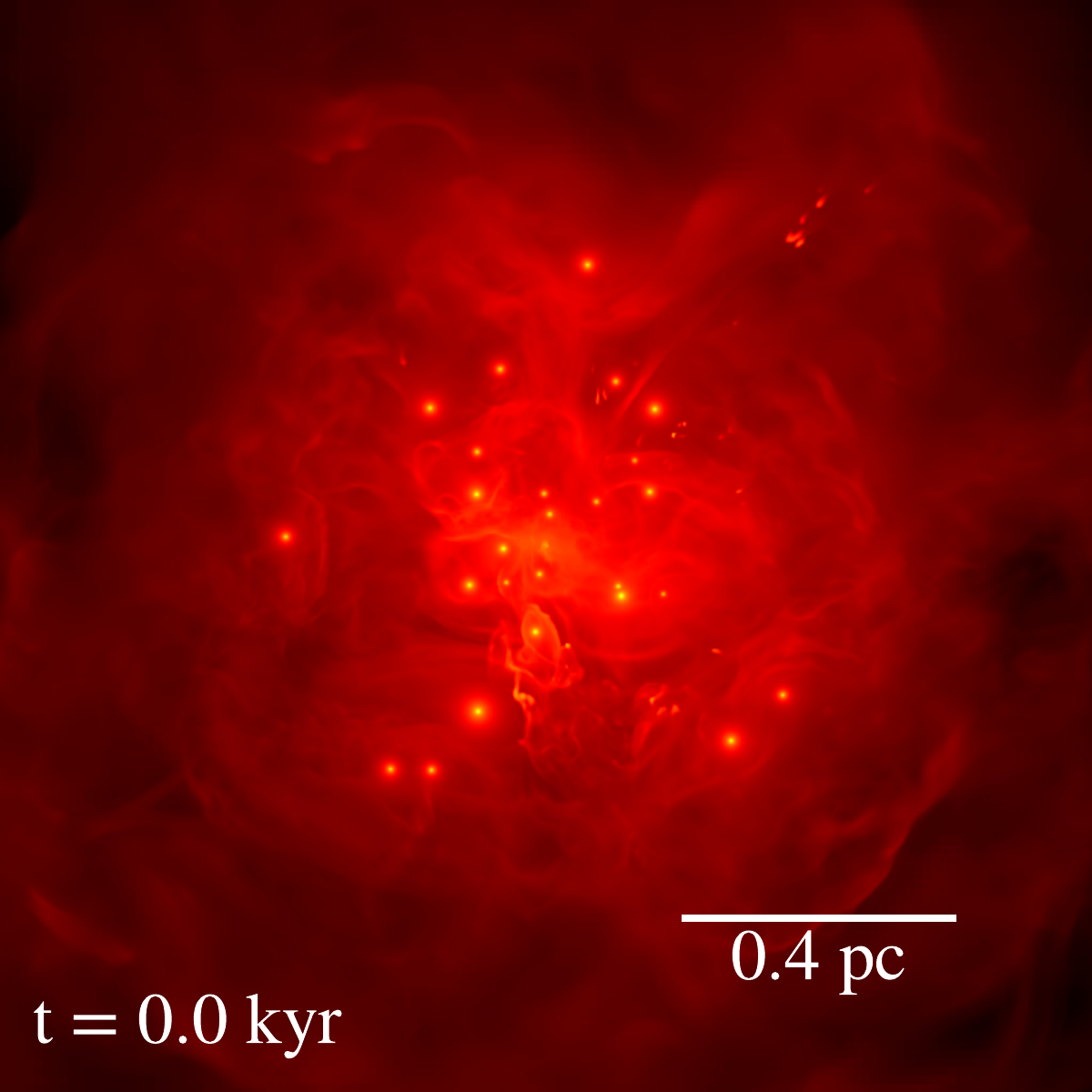}
            \includegraphics[width=0.325\linewidth]{plots/t0_hllc_z3.pdf}
            \includegraphics[width=0.325\linewidth]{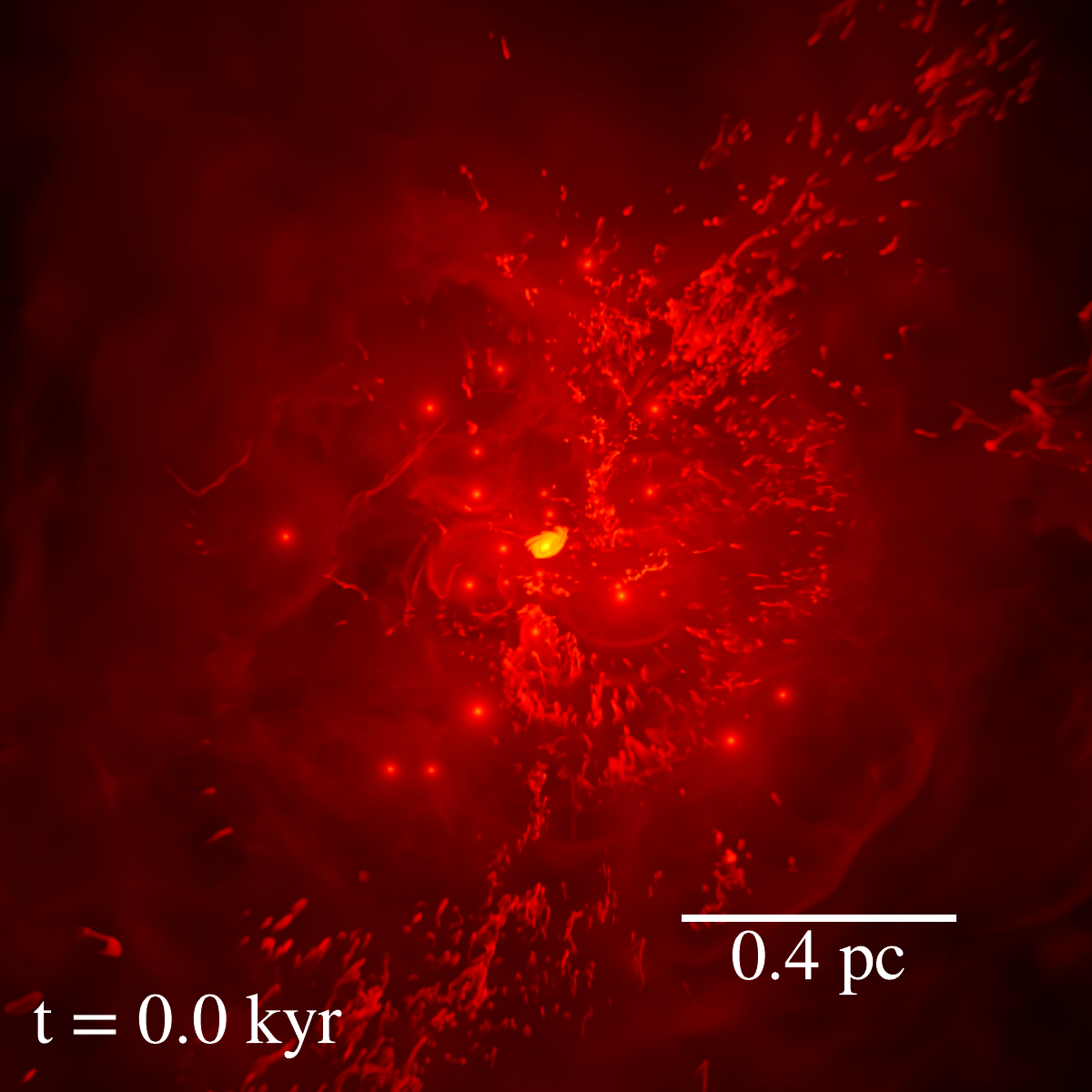}
            \caption{
            Comparison of the simulations with Solar composition varying the metallicity at two different simulation times.  
            The panels show projected density maps weighed by density along the $z$ axis, i.e. $\int \rho^2dz/\int\rho dz$, which is parallel to the line of sight. 
            Left-hand side, central, and right-hand side panels show the runs A1, A3, and A5, respectively. 
            All maps display the full computational domain.}
            \label{fig:solar}
        \end{figure*}

        Before investigating the models with the new implementation of the differential cooling we ran some experiments simply modifying the contribution of the metals to the total radiative cooling. 
        We simulated three models varying only the composition used: A1 used Solar composition, A3 used Solar composition with $3Z_{\odot}$, and A5 used Solar composition with $5Z_{\odot}$. 
        The density projected (weighted by density) maps of the final state of the simulations are shown in Figure~\ref{fig:solar}. 
        The left-hand side, central, and right-hand side panels display the runs A1, A3, and A5, respectively. 
        A higher metallicity in the plasma increases the radiative cooling directly. 
        As a result, more and denser filaments, clumps, and a denser central disc are obtained with increasing the metallicity. 
        On the contrary, if the metallicity is decreased to a Solar value we observe almost no overdensities, and definitively no disc formation. 
        We speculate that at $Z\approx 3Z_{\odot}$ there is a transition between the regimes of persistent quasi-steady accretion and disc formation, which would imply that small differences in the cooling function or numerical implementation can result in models reproducing either regime. 
        This may explain the differing results obtained by \cite{calderon2020} and \cite{ressler2020} with similar numerical and physical setups.

\begin{figure}
            \centering
            \includegraphics[width=\linewidth]{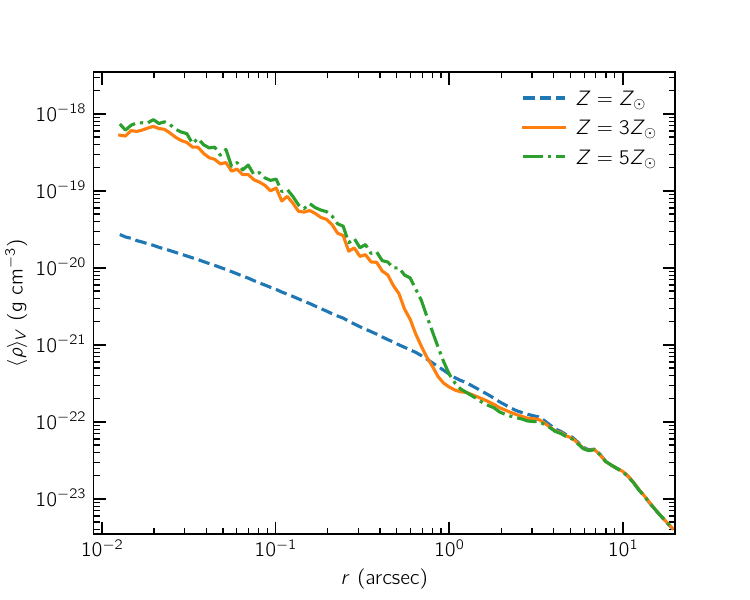}
            \includegraphics[width=\linewidth]{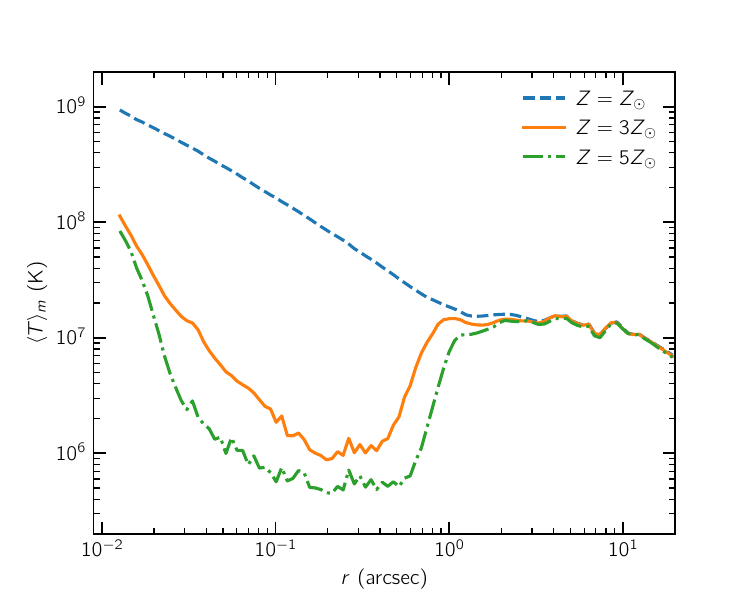}
            \caption{
            Radial profiles of time-averaged volume-weighted density (top) and mass-weighted temperature (bottom) over the last 500~yr of simulation time. 
            The models A1, A3, and A5 are shown in solid orange, dashed blue, and dashed-dotted green lines, respectively.
            }
            \label{fig:solar_prof}
        \end{figure}

        In order to perform a more quantitative comparison of these results Figure~\ref{fig:solar_prof} presents radial profiles of density (weighted by volume) and temperature (weighted by mass) on the top and bottom panels, respectively. 
        The dashed blue, solid orange, and blue dotted-dashed lines represent the models A1, A3, and A5, respectively. 
        The density profiles clearly shows the impact of cooling in the structure of the medium. 
        More efficient cooling, like in cases A3 and A5, results in an order of magnitude denser medium within 1\arcsec compared to the case with inefficient cooling, i.e. A1. 
        Notice that even higher metallicity in A5 extends the impact the denser region to slightly larger spatial scales. 
        The effect is similar in the temperature profile. 
        In A1, the temperature profile decays with $r$ within the central 1\arcsec. 
        But when the metallicity is high enough to make cooling efficient the temperature profile can decrease up to two orders of magnitude depending on the exact value. 

        In summary, these experiments allowed us to quantify the impact of radiative cooling by simply decreasing or increasing the radiative cooling influence. 
        However, since the chemical composition in reality is much more complicated and differs from Solar composition significantly these models should be interpreted only as general guides to assess the role of radiative cooling. 
        In the main text of this work, we devoted to explore more physically motivated mixtures that, although depend on more uncertain parameters are more robust when attempting to build a physical model that can be constrasted with observational data.

\end{appendix}
\end{document}